\documentclass[iop,apj,appendixfloats]{emulateapj}
\usepackage{apjfonts}
\usepackage{graphicx}
\usepackage{amsmath}
\usepackage{amssymb}
\usepackage{color}

\gdef\xx[#1]{\textcolor{red}{#1}}

\gdef\kms{km\,s$^{-1}$}
\gdef\msun{M$_{\odot}$}
\gdef\n1407{NGC\,1407}
\gdef\n1600{NGC\,1600}
\gdef\n2695{NGC\,2695}
\gdef\nai{Na\,{\sc i}\,$\lambda\lambda 8183,8195$\,\AA}
\gdef\nad{Na\,D\,$\lambda\lambda 5892,5898$\,\AA}
\gdef\feh{FeH\,$\lambda 9920$\,\AA}
\gdef\mg{Mg\,$\lambda 5177$\,\AA}

\lefthead{van Dokkum et al.}
\righthead{}
\slugcomment{Accepted for publication in the Astrophysical Journal}
\begin{document}

\title{The Stellar Initial Mass Function in Early-Type Galaxies
from Absorption Line Spectroscopy. III.
Radial Gradients}

\author{Pieter van Dokkum\altaffilmark{1}, Charlie Conroy\altaffilmark{2}, Alexa Villaume\altaffilmark{3}, Jean Brodie\altaffilmark{3},
Aaron J.\ Romanowsky\altaffilmark{4,5}
\vspace{8pt}}

\altaffiltext{1}
{Astronomy Department, Yale University, New Haven, CT 06511, USA}
\altaffiltext{2}
{Department of Astronomy, Harvard University, Cambridge,
MA 02138, USA}
\altaffiltext{3}
{Department of Astronomy and Astrophysics, University of California,
Santa Cruz, CA 95064, USA}
\altaffiltext{4}
{Department of Physics and Astronomy, San Jos\'e State
University, San Jose, CA 95192, USA}
\altaffiltext{5}
{University of California Observatories, 1156 High St., Santa Cruz, CA 95064,
USA} 

\begin{abstract}
There is good evidence that the centers of massive
early-type galaxies
have a bottom-heavy stellar initial mass function (IMF) compared to
the IMF of the Milky Way. Here we study the radial
variation of the IMF within
such galaxies, using a combination of high quality Keck spectroscopy
and a new suite of stellar population synthesis models that
cover a wide range in metallicity.
As in the previous studies in this series,
the models are fitted directly to the spectra and treat all elemental
abundance ratios as free parameters.
Using newly obtained
spectroscopy for six galaxies, including deep data extending to
$\sim 1R_{\rm e}$ for the galaxies NGC\,1407, NGC\,1600, and NGC\,2695,
we find that the IMF varies strongly with galactocentric radius.
For all six galaxies the IMF is bottom-heavy in the central regions,
with average mass-to-light ratio
``mismatch'' parameter
 $\alpha \equiv (M/L)/(M/L)_{\rm MW} \approx 2.5$ at $R=0$.
The IMF rapidly becomes more bottom-light with
increasing radius, flattening off near the Milky Way value
($\alpha \approx 1.1$) at $R>0.4R_{\rm e}$. 
A consequence is that the luminosity-weighted average IMF
depends on the measurement aperture: within
$R=R_{\rm e}$ we find
$\langle \alpha \rangle_L = 1.3-1.5$,
consistent with recent lensing and dynamical results from SLACS and
ATLAS$^{\rm 3D}$.
Our results are also consistent with several earlier studies that were based
on analyses of radial gradients of line indices. The observed
IMF gradients support
galaxy formation models in which the
central regions of massive galaxies had a different formation history
than their outer parts.
Finally, we make use of the high signal-to-noise central spectra of NGC\,1407
and NGC\,2695 to demonstrate how we can disentangle IMF effects and
abundance effects.

\end{abstract}


\section{Introduction}

There is strong evidence that the stellar
initial mass function (IMF) in the centers of massive
early-type galaxies is bottom-heavy with respect to the IMF in
the Milky Way disk. The evidence primarily
comes from three distinct observations.
First, recent stellar population synthesis (SPS)
modeling consistently indicates a
relatively
large contribution of low mass stars to the integrated light
(e.g., {Cenarro} {et~al.} 2003; {van Dokkum} \& {Conroy} 2010, 2011; {Conroy} \& {van Dokkum} 2012b; {Smith}, {Lucey}, \& {Carter} 2012; {Spiniello} {et~al.} 2012; {La Barbera} {et~al.} 2013; {La Barbera}, {Ferreras}, \&  {Vazdekis} 2015).
Second, the dynamics of
massive galaxies are better fitted when a {Salpeter} (1955) IMF is
assumed for the stellar component than
when a {Kroupa} (2001) or {Chabrier} (2003) IMF is used, as a Salpeter IMF
implies a total mass in stars that is higher by a factor of 1.6
(e.g., {Thomas} {et~al.} 2011; {Cappellari} {et~al.} 2012; {Dutton}, {Mendel}, \& {Simard} 2012). 
{Dutton} {et~al.} (2012) show that this also holds for very compact
galaxies, which are thought to have only a minor contribution from
dark matter within their effective readius.
Third, gravitational lensing studies show a behavior similar to
the dynamical studies, requiring a heavy, Salpeter-like IMF for the highest
mass lenses (e.g., {Treu} {et~al.} 2010; {Spiniello} {et~al.} 2011; {Sonnenfeld} {et~al.} 2015).

It is encouraging that all three methods are qualitatively consistent,
and that comparisons of different techniques generally show the
same trends.
{Conroy} {et~al.} (2013) 
show that dynamical and 
stellar population synthesis modeling of compact galaxies are consistent
with each other, and {Posacki} {et~al.} (2015) combine lensing with stellar
population modeling. {Lyubenova} {et~al.} (2016) find good agreement between
dynamical and stellar population measurements of galaxies in the CALIFA
survey, carefully controlling for systematic differences in the
methodology.
Taking these studies a step further,
{Spiniello} {et~al.} (2015a) combine lensing, dynamics, and
SPS modeling to constrain both the slope and the cutoff of the
IMF below 1\,\msun\ for a sample
of nine galaxies.

However, not all studies are in agreement,
and the question whether the IMF varies between galaxies
is not yet settled.
As shown by {Smith} (2014)
a direct comparison of published results of the same individual galaxies
using different techniques shows very large scatter.
In addition, a detailed study of several very nearby strong lenses
suggests tension between the lensing masses and the SPS-inferred ones
({Smith}, {Lucey}, \& {Conroy} 2015). Another example of possible tension is
the lack of variation in the number of X-ray binaries with
galaxy velocity dispersion ({Peacock} {et~al.} 2014), although this constrains
the IMF at high masses, not low masses.

It is likely that these disagreements, and
the large differences between some independent
measurements of the same 
galaxies (see {Smith} 2014), are due to a combination
of factors. First, it seems that there is considerable galaxy-to-galaxy
scatter in the IMF (e.g., {Conroy} \& {van Dokkum} 2012b; {Leier} {et~al.} 2016).
Furthermore,
it is certainly the case that the random errors
in all methods underestimate the true uncertainty
(see, e.g., {Tang} \& {Worthey} 2015). A striking illustration
of this  is Fig.\ 12 of {Conroy} \& {van Dokkum} (2012b), which shows that the
SPS-derived IMF can vary between Milky Way-like
and super-Salpeter for different
model assumptions and spectral fitting regions.\footnote{It should be
emphasized that these uncertainties mostly affect the overall normalization
of the IMF; in all panels of Fig.\ 12 in Conroy \& van Dokkum (2012b) there
is strong evidence for IMF variation between galaxies.}
Also, different techniques are sensitive to different stellar mass
ranges, as explored effectively in {Spiniello} {et~al.} (2015a). Whereas
lensing and dynamics measure the total mass, which includes stellar
remnants, dark matter, and gas,
SPS methods
are sensitive to the light of stars in specific mass ranges
(see Fig.\ 17 of Conroy \& van Dokkum 2012a, and Conroy
et al.\ 2017).

Another possible explanation for the variation between studies is that the
IMF may not only vary {\em between} galaxies but also {\em within} galaxies.
If this is the case, the use of different effective apertures will
introduce scatter even if the same methodology is applied to the same
objects. Also, for a given projected aperture,
lensing is sensitive to the mass in a cylinder,
dynamics to the mass in a sphere, and stellar population synthesis to the
projected light.
IMF gradients may be expected, because the velocity dispersion,
surface mass density, metallicity, age, and $\alpha$-enhancement all
change with radius (e.g., {Mehlert} {et~al.} 2003; {Kuntschner} {et~al.} 2010), and
the IMF may correlate with these parameters ({Conroy} \& {van Dokkum} 2012b; {Hopkins} 2013).
There are also reasons to expect an IMF gradient from the formation history
of massive galaxies. There
is evidence that the centers of many massive galaxies were assembled
in a short period of intense star formation at $z\gtrsim 2$
({Bezanson} {et~al.} 2009; {Oser} {et~al.} 2010; {Barro} {et~al.} 2013; {Nelson} {et~al.} 2014) that is distinct from
their later growth. The physical conditions inside these star forming cores
were very different from those in the Milky Way disk today
(e.g., {Zolotov} {et~al.} 2015; {van Dokkum} {et~al.} 2015; {Barro} {et~al.} 2016).

In the previous
papers in this series ({van Dokkum} \& {Conroy} 2012; {Conroy} \& {van Dokkum} 2012b) we used
an effective aperture of $R<R_{\rm e}/8$, where $R_{\rm e}$ is the
projected half-light radius. 
This small aperture was largely determined by the signal-to-noise ratio
(S/N) that is required to measure the IMF-sensitive \feh\
band ({Wing} \& {Ford} 1969). 
For comparison,
dynamical studies typically quote results at $\sim 0.5-1R_{\rm e}$.
In this paper we extend this analysis to larger radii, using newly obtained
data from the Keck\,I telescope in combination with an updated suite of
stellar population synthesis models.

This is not a new topic: studies of gradients in
IMF-sensitive spectral features  go back
at least to {Boroson} \& {Thompson} (1991), who found that the \nai\
doublet increases toward the centers of early-type galaxies.
The difficulty is that stellar abundances also change with radius;
{Boroson} \& {Thompson} (1991) could not clearly distinguish an IMF gradient from
a gradient in the sodium abundance (see also {Worthey}, {Ingermann}, \&  {Serven} 2011).
Recent studies have provided  superficially somewhat conflicting results,
although this may largely be due to differences in modeling techniques.
{Mart{\'{\i}}n-Navarro}  {et~al.} (2015a) find evidence for strong IMF gradients 
(from bottom-heavy in the center to Milky Way-like at large radii)
in two massive early-type galaxies, and a constant IMF in a low
mass galaxy. Similarly, {La Barbera} {et~al.} (2016) derive a strong IMF
gradient in a single massive elliptical galaxy using deep optical
and near-IR spectroscopy with the Very Large Telescope.
In apparent contrast, {McConnell}, {Lu}, \&  {Mann} (2016) suggest that the
observed line index gradients
of two massive early-type galaxies can be fully explained by
abundance variations, and {Zieleniewski} {et~al.} (2016) show that
a Milky Way IMF is consistent with their two-dimensional spectroscopy of
two of the three brightest galaxies in the Coma cluster.
Both studies emphasize that IMF effects are subtle, and that even if IMF
trends are present, abundance variations will likely dominate the
observed radial
changes in the strength of absorption lines.
Adopting a different approach,
{Davis} \& {McDermid} (2017) find significant variation in the IMF gradients
among seven galaxies using their molecular gas kinematics.


In this paper, we build on this previous work using optical
spectroscopy
extending to $1\,\mu$m obtained with the dual-beam
Low Resolution Imaging Spectrograph (LRIS; {Oke} {et~al.} 1995) on Keck.
We obtained high quality spectra out to $\gtrsim 1 R_{\rm e}$, thanks to a
custom long slit mask and the fact that we spent half of the observing
time on empty sky fields. 
Following the methods we used in previous papers in this series, and
in contrast to other studies of IMF gradients,
we fit state of the art models directly to the spectra rather than
to indices.




\begin{figure*}[htbp]
  \begin{center}
  \includegraphics[width=0.7\linewidth]{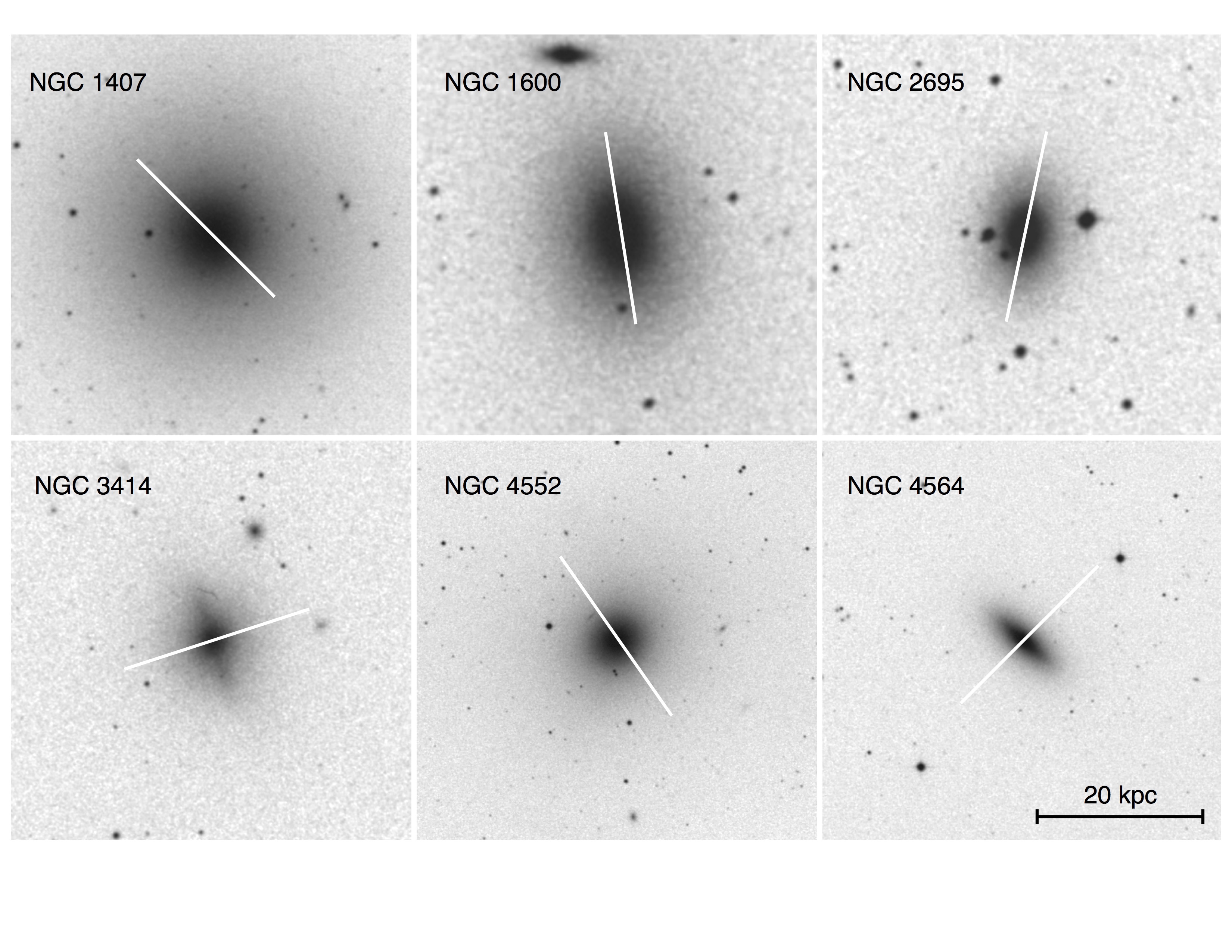}
  \end{center}
\vspace{-1.3cm}
    \caption{
DSS images of the six newly observed galaxies, scaled to a common distance.
White lines indicate the orientation of the LRIS slit. We
obtained deep observations along the major axis
for the three galaxies in
the top panels (NGC\,1407, NGC\,1600, and NGC\,2695), interspersed with
independent off-target sky exposures. The galaxies in the bottom panels
have shallower observations, along the minor axis.
}
\label{galimages.fig}
\end{figure*}

\section{Observations}

We obtained spatially-resolved spectroscopy of six early-type
galaxies  on December 19--20 2014, using LRIS
on the Keck\,I
telescope. For three of the galaxies we obtained deep
data with a special
long slit, and interspersed the science images 
with blank sky exposures of equal integration time.

\begin{deluxetable}{cccccc}
\tablecaption{Galaxy Sample\label{sample.tab}}
\tabletypesize{\footnotesize}
\tablehead{\colhead{Id} & \colhead{$D$\tablenotemark{a}} &
\colhead{$R_{\rm e}$\tablenotemark{b}} & \colhead{$\sigma_0$\tablenotemark{c}}
& \colhead{PA$_{\rm slit}$} & \colhead{$t_{\rm exp}$\tablenotemark{d}}
\\
\colhead{} & \colhead{[Mpc]} & \colhead{[$'$]} & \colhead{[km\,s$^{-1}$]} &
\colhead{} & \colhead{[s]}}
\startdata
NGC\,1407 & 23.3 & 1.26 & 292 & maj & 7800 \\
NGC\,1600 & 45.8 & 1.12 & 340 & maj & 9000 \\
NGC\,2695 & 35.3 & 0.42 & 229 & maj & 3000 \\
NGC\,3414 & 23.5 & 0.49 & 240 & min & 1800 \\
NGC\,4552 & 16.0 & 0.52 & 264 & min & 1800 \\
NGC\,4564 & 17.0 & 0.24 & 163 & min & 1800
\enddata
\tablenotetext{a}{Averages from the NASA/IPAC Extragalactic Database.}
\tablenotetext{b}{Effective radii along the direction of the slit. For
NGC\,1407, NGC\,1600, and NGC\,2695 this is the major axis; for the
other galaxies this is the minor axis.}
\tablenotetext{c}{Measured values, uncorrected for seeing, from
the central spectra in this paper.}
\tablenotetext{d}{Total on-target exposure time. For the first three
galaxies, an equal amount of time was spent off-target.}
\end{deluxetable}


\subsection{Sample}

The three primary targets of our study  are the giant elliptical galaxies
NGC\,1407 and  NGC\,1600 and the S0 galaxy NGC\,2695.
NGC\,1407 is the dominant member of the eponymous NGC\,1407 group
at 24\,Mpc, which
also includes the elliptical galaxy NGC\,1400
({Brough} {et~al.} 2006; {Romanowsky} {et~al.} 2009). NGC\,1600, at
$\sim 50$\,Mpc,
is generally considered
an isolated elliptical, with an extensive system of satellite galaxies
({Smith} {et~al.} 2008). NGC\,2695 is one of the brightest galaxies in a
group that also contains the elliptical galaxy NGC\,2699.
Distances are averages taken from NED.\footnote{https://ned.ipac.caltech.edu/}
These distances come from a variety of sources and may
be uncertain, but as we express nearly all
our results as an IMF ratio ($(M/L)/(M/L)_{\rm MW}$)
versus a radius ratio ($R/R_{\rm e}$)
they are independent of the absolute distances.

NGC\,1407 and NGC\,1600 are very similar:
both are slowly rotating, very large and very massive galaxies.
Their central velocity dispersions are
292\,\kms\ and 340\,\kms\ respecively (see \S\,\ref{starpops.sec}).
They likely have massive central black holes,
with the black hole in NGC\,1600 recently
claimed to be among the most massive
in the local Universe ({Thomas} {et~al.} 2016).
Their major axis half-light radii are $1\farcm 26$
and $1\farcm 12$ ({Li} {et~al.} 2011), corresponding to 8.8\,kpc and
16\,kpc.\footnote{Note that the effective radii of massive galaxies
such as these are somewhat uncertain; see, e.g., Bernardi et al.\ 2014.}
NGC\,1407 is nearly round, with an ellipticity of $\epsilon = 0.04$. NGC\,1600
has $\epsilon = 0.39$. 
NGC\,2695 is a rotating S0 galaxy. It was
chosen largely because of its availability at the end of the night.


In addition to these galaxies we observed three others, with
the slit oriented along the minor axis:
NGC\,3414, NGC\,4552, and NGC\,4564. These galaxies were
selected because of their relatively high [Na/Fe] values
in data we had obtained previously.\footnote{These earlier,
as yet largely unpublished, data did not cover the Na\,D
doublet.}
Exposure times were shorter
for these objects (although still substantially longer
than in van Dokkum \& Conroy 2012),
and we cannot measure their IMF gradients to the same
distance as for the primary galaxies.
The six galaxies are shown in Fig.\ \ref{galimages.fig}.
The images
were scaled to the same distance so their spatial extent can be compared
directly. Basic information for the galaxies is provided in Table
\ref{sample.tab}. The semi-major axis
effective radii of NGC\,1407 and NGC\,1600 were
taken directly from {Li} {et~al.} (2011). The effective radii of the other
galaxies were determined from the circularized values and 
ellipticities listed in {Cappellari} {et~al.} (2007).

\subsection{Methodology}

LRIS is a dual beam optical spectrograph, providing simultaneous
high sensitivity observations from the far blue to the far red. 
The beams were split with the D680 dichroic, which has a 50\,\% reflectance
wavelength of 6640\,\AA\ and a 50\,\% transmission wavelength
of 6800\,\AA.
In the blue arm, we used the 300\,l\,mm$^{-1}$ grism blazed at 5000\,\AA.
This is a departure from the strategy of
{van Dokkum} \& {Conroy} (2012) and  {McConnell} {et~al.} (2016), who
used the higher resolution 600\,l\,mm$^{-1}$ grism. The 300\,l\,mm$^{-1}$
grism covers the full spectral range from the atmospheric cutoff to the
dichroic. The 600\,l\,mm$^{-1}$ grism covers wavelengths $\lambda \lesssim
5600$\,\AA, missing the \nad\ doublet. As shown in Fig.\ 12 of
{Conroy} \& {van Dokkum} (2012a) (and in \S\,\ref{imftest.sec}),
the Na\,D line is important for distinguishing IMF effects
from variations in the sodium abundance. 
In the red arm the 600\,l\,mm$^{-1}$ gold-coated grating blazed
at 10000\,\AA\ was used, as in previous work.
The LRIS-red detector has fully depleted, high-resistivity CCDs
(see {Rockosi} {et~al.} 2010). These devices are sensitive to wavelengths
$>1\,\mu$m, and have no appreciable fringing. This makes it possible
to do accurate, sub-percent spectroscopy in the far red.

The standard long slit of LRIS covers only approximately half of the
field of view. Furthermore, in both the blue and the red beams
the middle of the slit falls in the gap between the two mosaiced
detectors. As a result the usable contiguous length of the slit on
a single chip is
only $\approx 90\arcsec$. We designed a custom slit mask, comprised of
a $0\farcs 7$ wide, $\approx 290\arcsec$ long
slit that is broken into four pieces to ensure mechanical
stability. This slit
is approximately twice as long as the standard long slit.
It has three gaps; the central one coincides with the detector gap.

For the three primary targets we used the following observing strategy.
We used the ``special'' long slit that we designed, aligned with the
major axis of the galaxy. 
The red side data were binned on-chip by
a factor of two to reduce the read-out time, providing
pixel scales of
$0\farcs 27$ in the red and $0\farcs 135$ in the blue. 
We  obtained a series of 600\,s exposures,
alternating on-target exposures with off-target exposures. These off-target
exposures are used in the reduction to enable very accurate sky subtraction
over the entire spatial range of interest. They
are $\sim 15\arcmin$ removed from the galaxies, and
carefully chosen so that no
stars or other contaminating objects fall in the slit. 
For each object an
equal number of on-target and off-target exposures was obtained. This
strategy is similar to that of {Kelson} {et~al.} (2002), who used LRIS to study
the kinematics of a brightest cluster galaxy to large radii.
{McConnell} {et~al.} (2016)
also took off-target exposures but less frequently (one sky exposure
for every 2--4 science exposures). 
The total on-target exposure times were 7800\,s, 9000\,s, and 3000\,s for
NGC\,1407, NGC\,1600, and NGC\,2695 respectively.

The other three galaxies were observed in a classical way, dispensing with the
off-target exposures. The slit was aligned with the minor axis,
to facilitate standard sky subtraction techniques. Three 600\,s exposures
were obtained for each galaxy, moving the telescope
along the slit by $30\arcsec$
in between exposures. The total exposure time was therefore 1800\,s
for each object.
The slit positions are indicated on Fig.\ \ref{galimages.fig} for all six
galaxies.

\subsection{Data Reduction}

\subsubsection{Two-dimensional Sky Subtraction}

The data reduction was done with a custom pipeline written
in the Python programming environment. We focus here on the reduction
of the three primary targets, as the analysis of the other galaxies
largely follows that described in {van Dokkum} \& {Conroy} (2012).
The first step in the reduction is to subtract a blank sky frame from
each of the science exposures.
The data were divided into sets
of three exposures, consisting of a science exposure and the two
adjacent blank sky exposures.
For each science exposure,
a two-dimensional sky frame was created by averaging the adjacent
sky exposures. This produces a good match to the sky in the science
exposure if changes in the sky line intensities were linear over the
$\sim 30$\,min that elapsed during the three exposures. To account
for non-linearity, the averaged sky frame was scaled to match the
science frame by measuring the fluxes of the brightest sky lines
in both images. This scaling was typically $\lesssim 1$\,\%, and only
important for data taken
in the evening or morning when approaching 18 degree twilight.

\subsubsection{Wavelength Calibration and Instrumental Resolution}

Next, the spectra were wavelength calibrated. The initial solution is
based on arc line exposures of the custom long slit taken in the afternoon.
We used all lamps. The cadmium, zinc, and mercury
lamps are the main calibrators in the blue
and the neon and argon lamps have many lines in the red.
In the standard line
list\footnote{https://www2.keck.hawaii.edu/inst/lris/txt/all\_line\_list.txt}
there is a rather large gap in wavelength coverage between the Ar lines
at 9787\,\AA\ and 10473\,\AA, a wavelength regime that contains the
\feh\ band. Fortunately
there is a faint Ar line at 10054.81\,\AA, and
we added this line to the list.

In both the blue and
the red, the detector consists of two chips with two amplifiers
each. The data from
each of the four amplifiers (two chips, and two amplifiers per chip)
were fitted separately.
In the red, a polynomial of 6$^{\rm th}$
order in the spatial direction
and 5$^{\rm th}$
order in the wavelength direction gives
an rms scatter of $\approx 0.08$\,\AA\ for each of the segments.
The full observed wavelength range is approximately
7450\,\AA\ -- 10750\,\AA. As a check on the wavelength calibration in
the far red we collapsed the 2D arc spectra in the spatial direction
and measured the location of the very weak $\lambda 10335.55$
Ar line. The measured
wavelength is 10335.5, which demonstrates that the polynomial fit
accurately captures the transformation from pixel coordinates to
wavelengths in this regime. In the blue, a polynomial of
6$^{\rm th}$ order
in both the spatial direction and the wavelength direction
gives an rms scatter of $\approx 0.15$\,\AA.
LRIS has considerable flexure, and the exact wavelength solution is a complex
function of the pointing of the telescope and other factors. For each
individual science exposure we applied
a zero-order offset to the high order arc line solutions.
The applied offsets are medians of the offsets calculated from multiple sky
emission lines. The offsets are typically $\sim 3$\,\AA\ in the blue and
$\sim 1$\,\AA\ in the red, with the exact effect
dependent on the field and time of night, and with
no obvious residual wavelength dependence.

\begin{figure}[htbp]
  \begin{center}
  \includegraphics[width=1.0\linewidth]{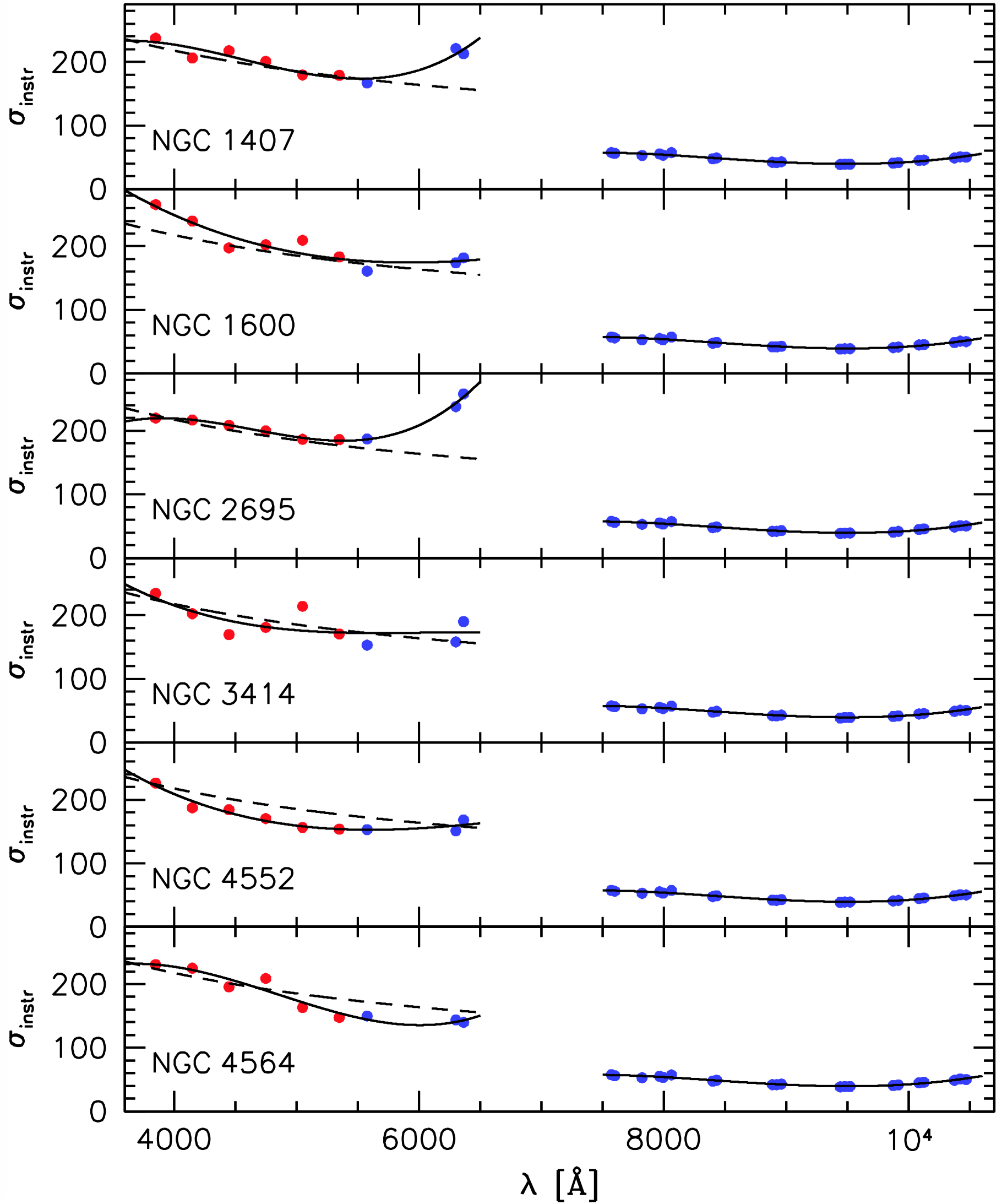}
  \end{center}
\vspace{-0.3cm}
    \caption{
Instrumental resolution (in \kms)
as a function of wavelength, as determined from
sky emission lines (blue) and fits to the spectra (red; see text).
Solid curves are fits to the data.
We use the same functional form for all galaxies in the red,
but due to focus variations we use a custom fit for each
of the six galaxies in the blue.
The blue resolution as determined
from arc lamps is shown by the dashed line, for reference.
}
\label{resol.fig}
\end{figure}

\begin{figure}[htbp]
  \begin{center}
\hspace{-0.5cm}
  \includegraphics[width=1.055\linewidth]{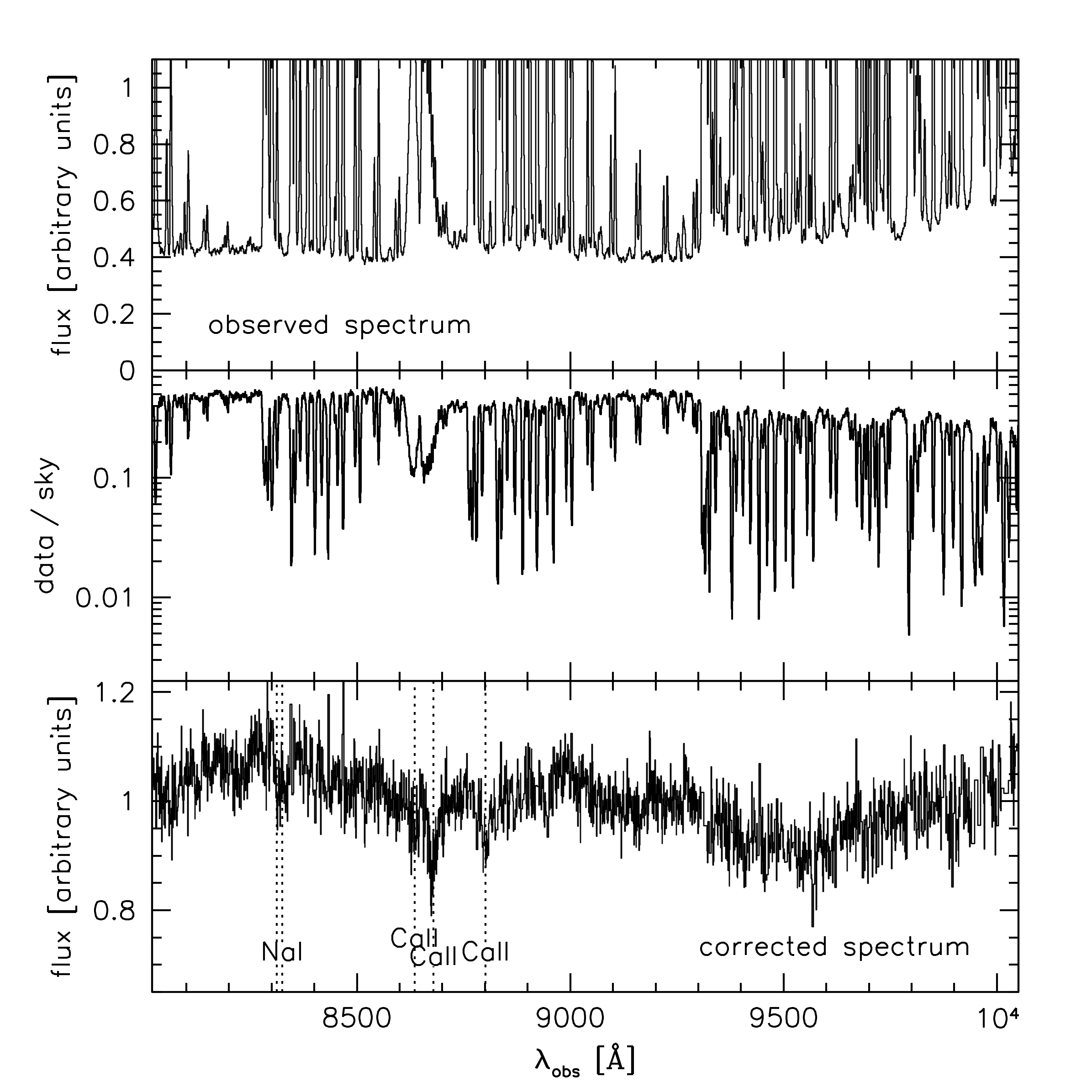}
  \end{center}
\vspace{-0.4cm}
    \caption{
Red-side spectrum near $r=65\arcsec$, or $1.0 R_{\rm e}$, for NGC\,1600.
The top panel shows the observed spectrum (the average of 28 rows).
The middle
panel is the ratio between the galaxy spectrum and the
sky. At $1R_{\rm e}$,
the galaxy flux is only 1\,\% -- 10\,\% of the
sky emission in the far red.
The bottom panel shows the sky subtracted spectrum, with prominent
spectral features marked.
}
\label{demo.fig}
\end{figure}

\begin{figure*}[htbp]
  \begin{center}
  \includegraphics[width=0.9\linewidth]{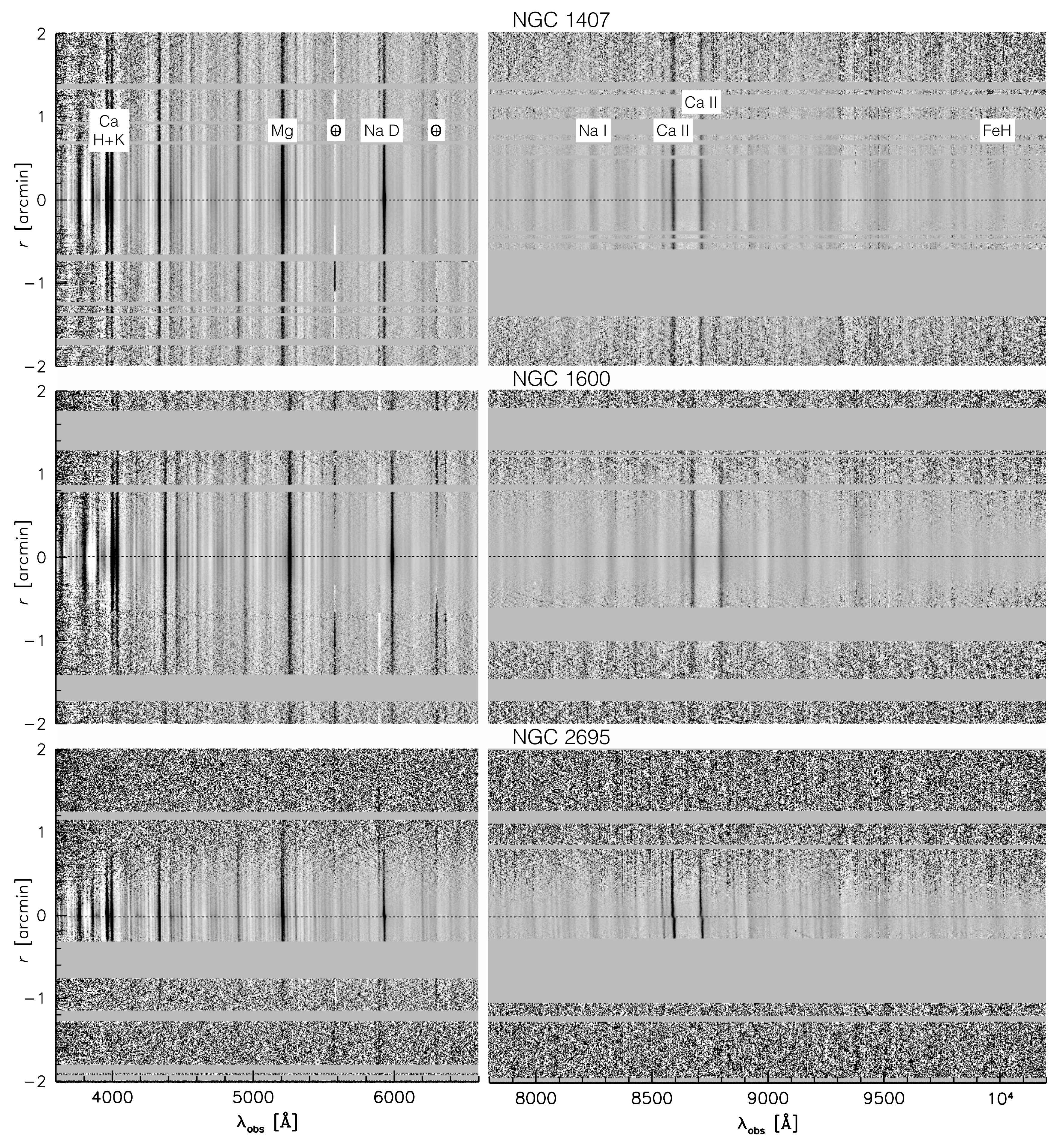}
  \end{center}
\vspace{-0.3cm}
    \caption{
Two-dimensional major axis
spectra for the three galaxies that were obtained with
our on/off observing strategy. The spectra were divided by a polynomial
in the wavelength direction to highlight the absorption features.
Grey horizontal bars indicate masked regions; these are
contaminating objects as well as the chip gap.
The blue spectra have a resolution that
ranges from $\sigma_{\rm instr}\approx 250$\,\kms\ to $\approx 150$\,\kms.
The red spectra have a much higher resolution of $\sigma_{\rm instr}
\approx 40$\,\kms, and velocity and velocity dispersion gradients
can be seen by eye.
Prominent spectral features are marked, as well as two strong night sky
emission lines.
}
\label{2dspec.fig}
\end{figure*}

For the stellar population fitting it is important to accurately
measure the instrumental resolution. 
In the red $\sigma_{\rm instr}\approx 40$\,\kms, much smaller than the
velocity dispersions of the galaxies, and it can be measured accurately
from sky emission lines (see Fig.\ \ref{resol.fig}).
In the blue the situation is more complex, and requires careful
treatment. First, due to our choice of a low resolution grism that
covers Na\,D, $\sigma_{\rm instr}=150-250$\,\kms, comparable
to the velocity dispersions of the galaxies. Second, there are no
sky emission lines that can be used blueward of the
$\lambda 5578$\,[O\,{\sc i}] line. Third, we find that there are
significant focus variations in the blue, particularly redward of
$5500$\,\AA. The variations seem random and
may be caused by temperature changes during the nights.

We measured the instrumental resolution in the blue in the following
way. All six galaxies were previously observed with the higher
resolution 600\,l\,mm$^{-1}$
grism (in 2012; see Conroy et al., in preparation). The instrumental
resolution of those data varies between $\sigma_{\rm instr,2012}
\approx 100$\,\kms\ at
3700\,\AA\ to $\sigma_{\rm instr,2012}
\approx 70$\,\kms\ at 5500\,\AA.
Using the same extraction aperture, we fit the 
new spectra with the high resolution 2012 data in six
wavelength intervals between 3700\,\AA\ and 5500\,\AA, with the
velocity dispersion $\sigma_{\rm fit}$
as the only free parameter.
The instrumental resolution of the new
data is then $\sigma_{\rm instr,2014}=
(\sigma_{\rm instr,2012}^2 + \sigma_{\rm fit}^2)^{1/2}$
in each wavelength interval. The results are shown by the red points
in Fig.\ \ref{resol.fig}. For each galaxy the red points connect
smoothly to the $\lambda>5500$\,\AA\ blue points,
demonstrating that our methodology is consistent with direct measurements
from sky emission lines. Solid lines show the third-order polynomials that
are used in the stellar population modeling.


\subsubsection{Residual Sky Subtraction, Image Combination, and
$s-$Distortion Correction}

The wavelength-calibrated individual amplifiers of each science exposure
were placed in a common 2D image, taking the detector gap into account.
Next, a zero-order residual sky subtraction was performed to account for
small differences in wavelength calibration and sky line intensity between
the science exposures and the adjacent sky frames. This is particularly
important for the broad O$_2$\,($0-1$) band at $\approx 8650$\,\AA, which
is independent of the OH lines and varies on short time scales.
To do this,
a small region at the bottom of the frame was used to measure the residual
sky spectrum in each science exposure.
As the center of the galaxy was placed on the top detector away from
the chip gap (the standard slit pointing origin), this region is
at a radius of approximately $3'$ from the center. We verified that the
galaxy flux at this radius is sufficiently low that its subtraction has
a negligible effect on the analysis in this paper.

The individual science frames were combined, scaling by the
collapsed galaxy flux and rejecting high and low pixels. The
distortion in the spatial direction (the 
$s-$distortion) was determined by measuring the central position
of the galaxy as a function of wavelength and fitting these positions
with a 3$^{\rm rd}$
order polynomial. The spectra were shifted, ensuring that
the center of the galaxy falls
on the center of a pixel in the corrected frame.

\subsection{Atmospheric Transmission and Response Function}

The corrections for telluric absorption and the instrument response
function follow the same procedures as described in detail in
{van Dokkum} \& {Conroy} (2012). Briefly, a theoretical atmospheric
transmission spectrum was fitted to the
observed central spectrum of the galaxy in
the wavelength interval 9300\,\AA\ -- 9700\,\AA, where there are
many strong H$_2$O lines. The fit also includes a polynomial to account
for the variation in the galaxy spectrum in this spectral range.
The fits converge quickly and provide a near-perfect removal of the
telluric absorption lines. We note that for the galaxies in this paper
there is no ambiguity, as the telluric lines are a factor of $\sim 6$
narrower than the galaxy absorption lines.
The spectra were corrected for the instrument response using observations
of the white dwarf Feige\,110. Special care was taken to correct for the
broad hydrogen Paschen lines in the observed white dwarf spectrum.


A graphic illustration of the 
reduction in the outermost spectral bins
is 
shown in Fig.\ \ref{demo.fig}. Accurate modeling of the sky
is critical, as the galaxy flux is only 1\,\% -- 10\,\%
of the sky emission in the far red. The reduced 2D spectra
of the three primary galaxies are shown in Fig.\ \ref{2dspec.fig}.
The spectra were divided by a polynomial in the wavelength direction
to reduce the dynamic range and highlight the absorption lines
at all radii.
With a few isolated exceptions (such as the strong
$\lambda$5578\,[O\,{\sc i}]
line) the spectra are very clean with no obvious systematic
issues.

\section{Fitting}

\subsection{Extracted Spectra as a Function of Radius}

For all six galaxies
we extracted one-dimensional (1D) spectra from the two-dimensional
(2D) reduced spectra. The apertures are defined in units of
binned ($0\farcs 27$) pixels. With the exception of the inner
apertures they are spaced quadratically, following the relation
$r_a = 3\times i^2$, with $i$ an integer and $r_a$ the
aperture radius in
pixels. This scaling is a compromise between having a sufficiently
fine sampling of the full radial range and maximizing the S/N ratio
in each bin.
The central aperture is 3 pixels ($0\farcs 81$),
corresponding to the approximate seeing. The galaxy spectrum
at each radius is defined as the sum of all image rows between
$r_a(i-1)$ and $r_a(i)$, not including rows that were masked
because of missing data (due to the chip gap) or because
of contaminating objects.

For each aperture in each galaxy we
calculated the luminosity-weighted mean radius $R$, properly
taking masked rows into account. Except for the central
aperture ($r=0$) each radius occurs twice, as spectra
are independently extracted from each side of the galaxy.
Because the weighting and masking are not the same on each
side of the galaxy, the positive and negative distances from
the center are not identical.
These luminosity-weighted radii  are
the ones that are used in the remainder of this
paper.

\begin{figure*}[htbp]
  \begin{center}
  \includegraphics[width=0.75\linewidth]{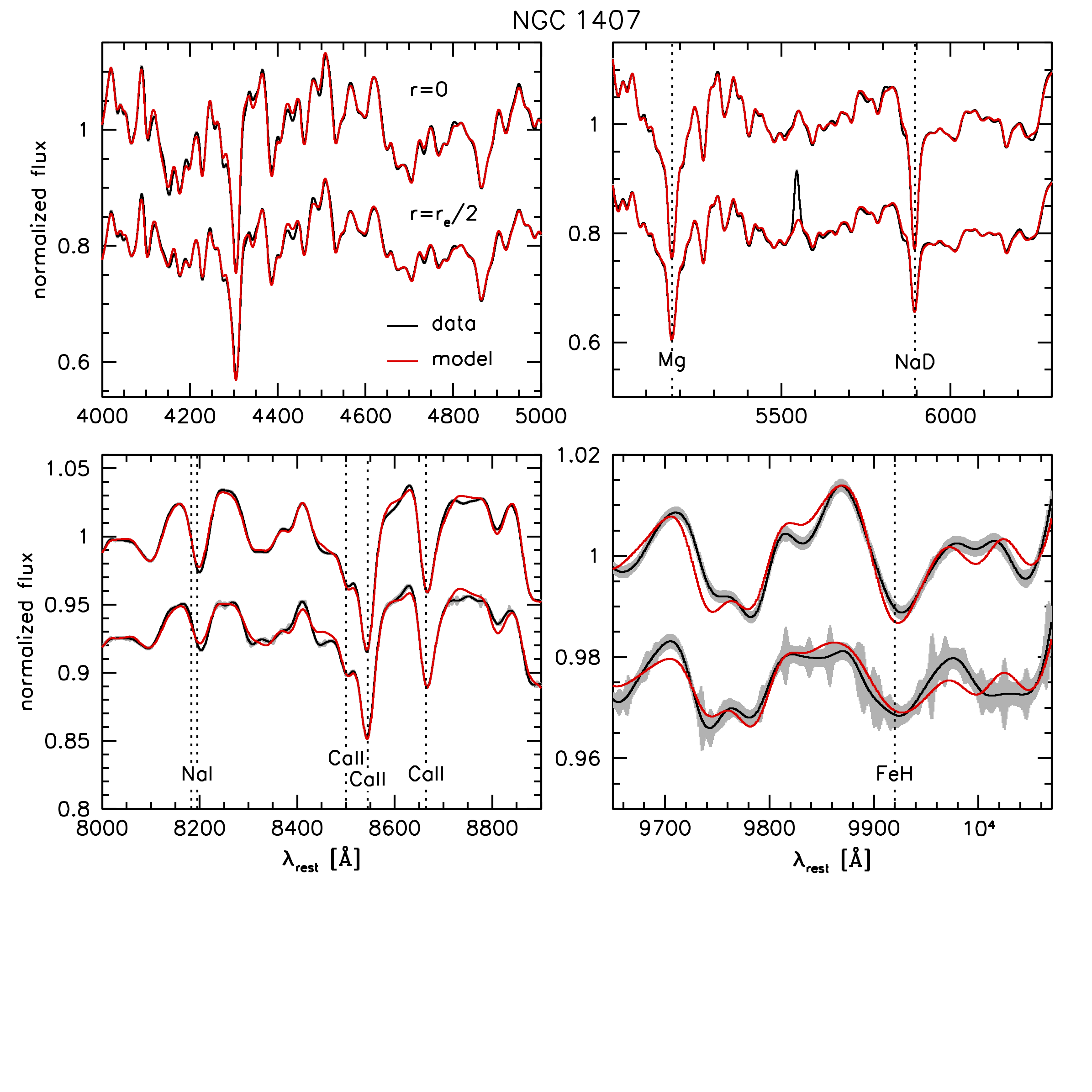}
  \end{center}
\vspace{-3cm}
    \caption{
Observed spectra (black) and best-fitting stellar population
synthesis models (red) for NGC\,1407, at $R=0$ (top) and
$R=R_{\rm e}/2$ (bottom).  Note that the \nai\ feature appears
offset to the red because it is blended
with a TiO bandhead at 8205\,\AA.
}
\label{full_spectra1.fig}
\end{figure*}

\begin{figure*}[htbp]
  \begin{center}
  \includegraphics[width=0.75\linewidth]{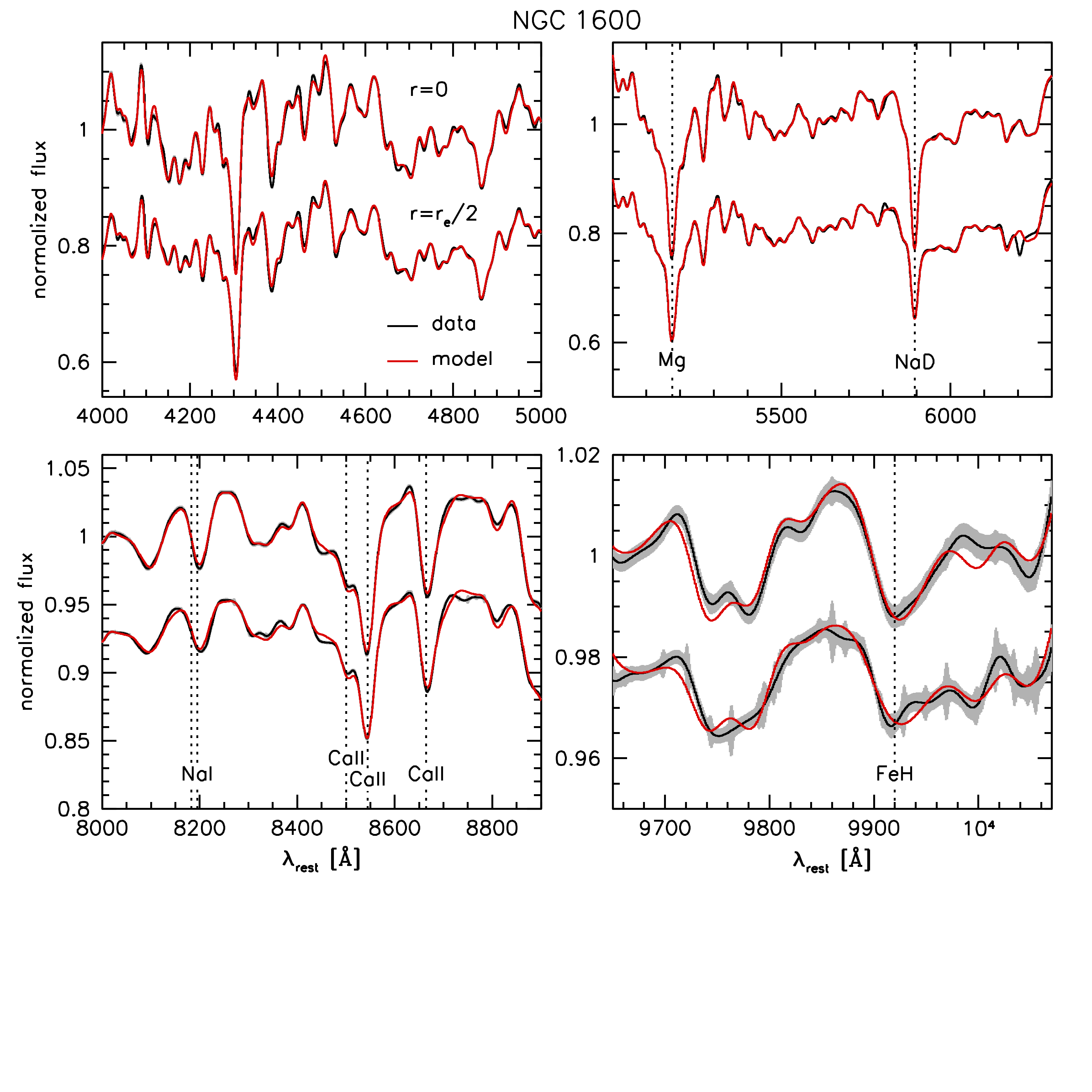}
  \end{center}
\vspace{-3cm}
    \caption{
Observed spectra (black) and best-fitting stellar population
synthesis models (red) for NGC\,1600, at $R=0$ (top) and
$R=R_{\rm e}/2$ (bottom). 
}
\label{full_spectra2.fig}
\end{figure*}

\begin{figure*}[htbp]
  \begin{center}
  \includegraphics[width=0.75\linewidth]{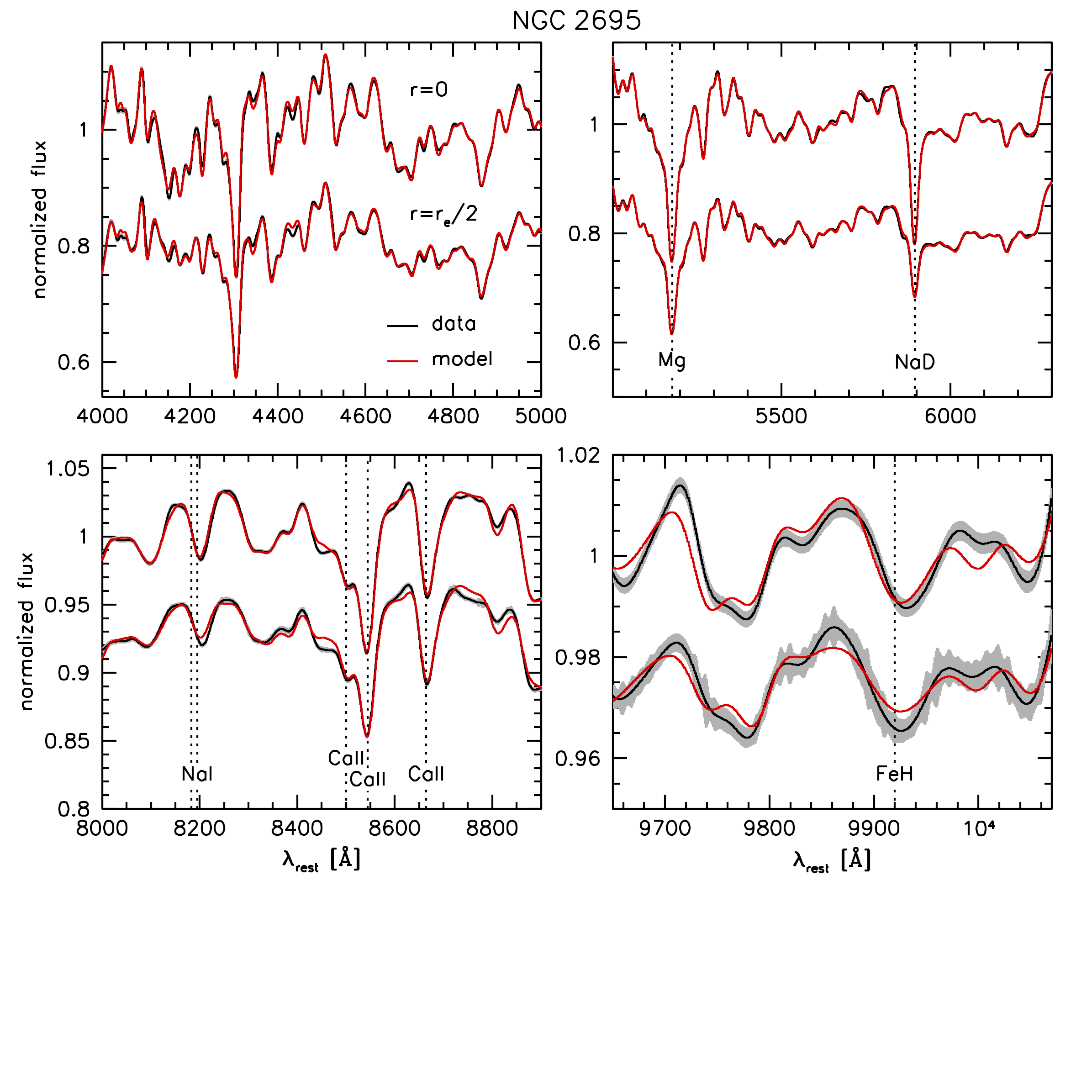}
  \end{center}
\vspace{-3cm}
    \caption{
Observed spectra (black) and best-fitting stellar population
synthesis models (red) for NGC\,2695, at $R=0$ (top) and
$R=R_{\rm e}/2$ (bottom).
}
\label{full_spectra3.fig}
\end{figure*}

\begin{figure*}[htbp]
  \begin{center}
  \includegraphics[width=0.65\linewidth]{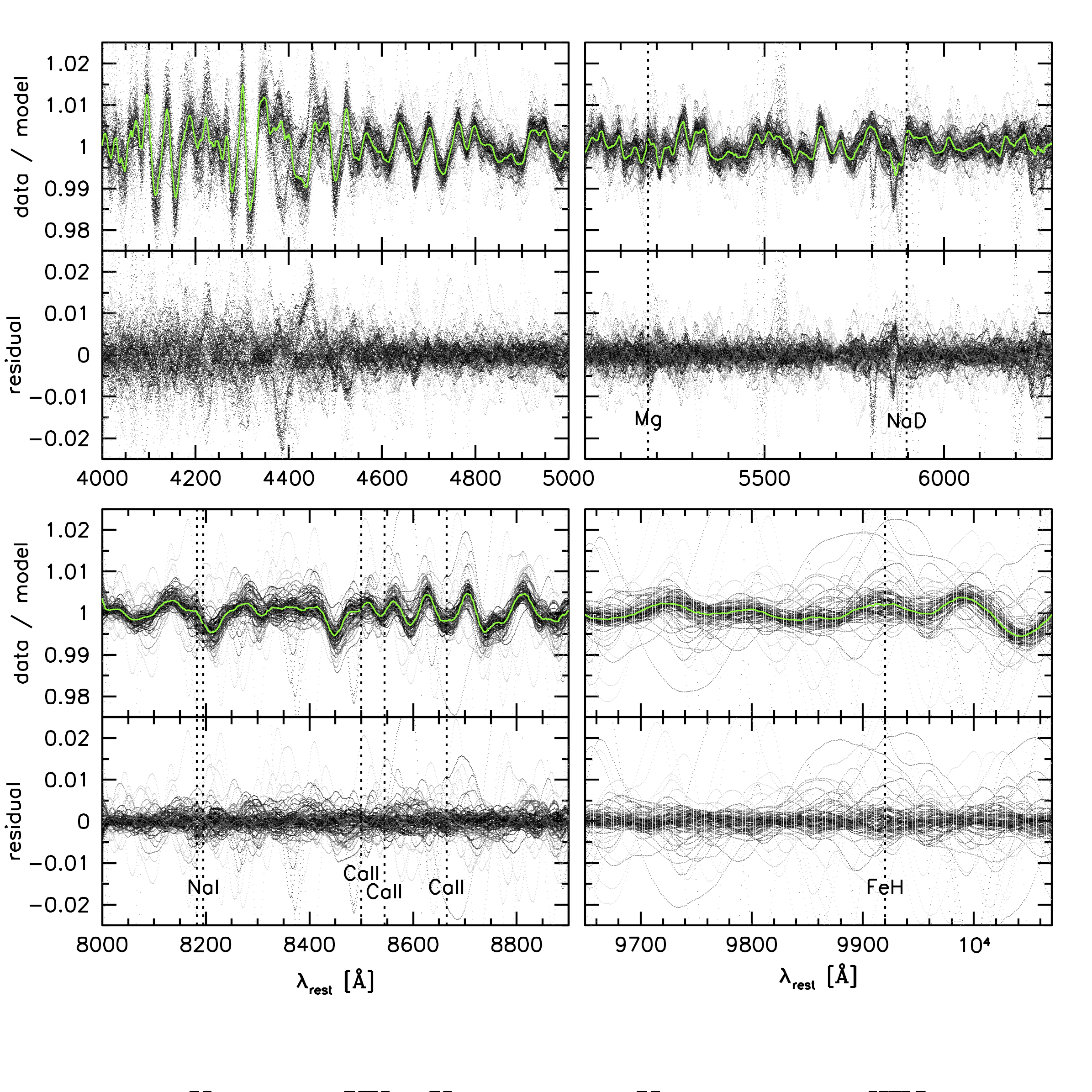}
  \end{center}
\vspace{-1.3cm}
    \caption{
Residuals from the fits, for all radial bins and all six
galaxies. Spectra of higher S/N ratio are displayed darker.
The green line in the upper panels is the median residual.
The residuals are nearly identical for all spectra,
and likely represent systematic errors in the models that
are independent of age, metallicity, and the IMF.
The
lower panels show the residual after subtracting the green line.
}
\label{res.fig}
\end{figure*}

\begin{figure*}[htbp]
  \begin{center}
  \includegraphics[width=0.75\linewidth]{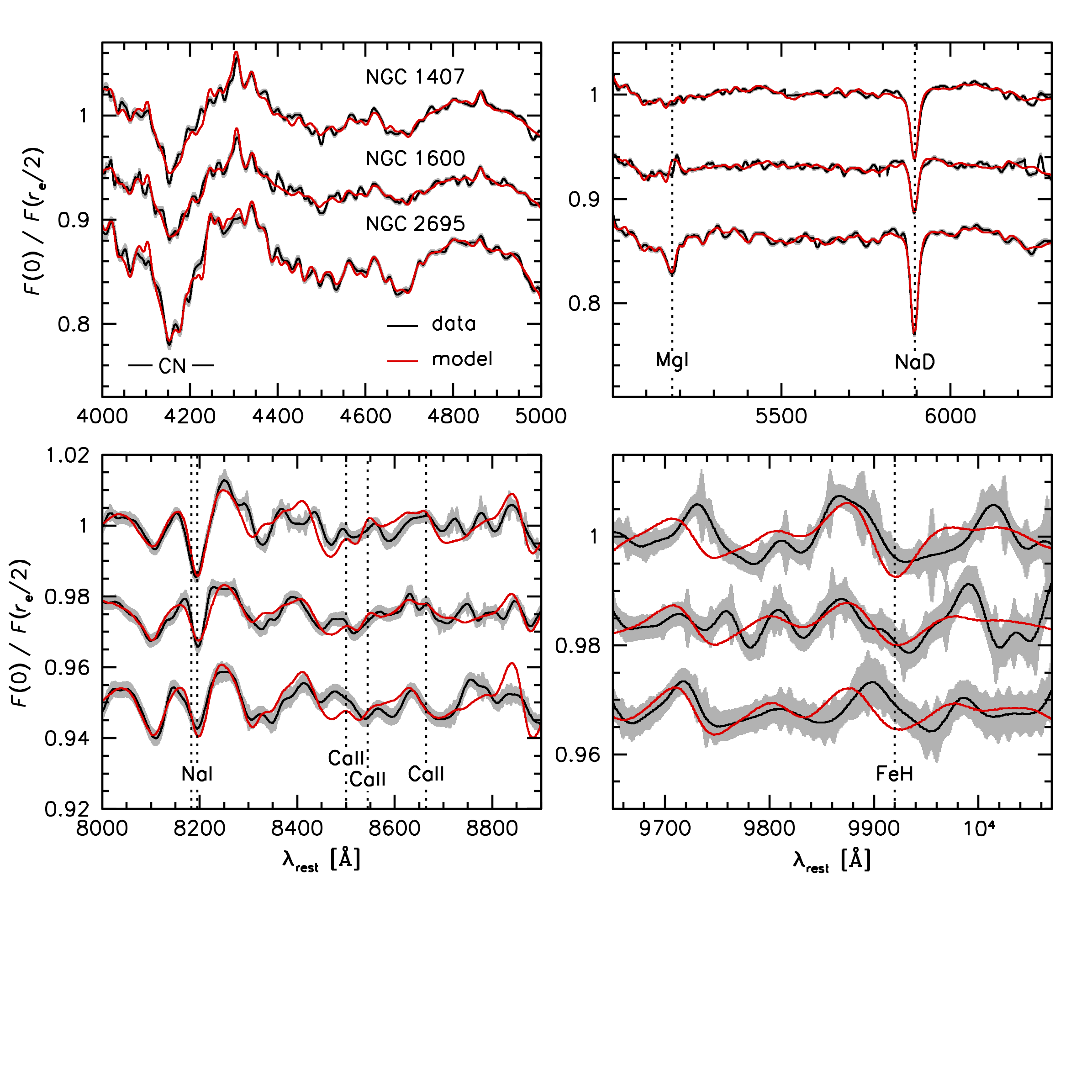}
  \end{center}
\vspace{-3cm}
    \caption{
Ratio of central spectrum to the spectrum at $R\sim R_{\rm e}/2$,
for the three primary galaxies. The data are shown in black
and the best-fitting models in red. The spectra of NGC\,1600
and NGC\,2695 were vertically offset for clarity. The ratio spectra
show significant features throughout the spectral range,
and these features are very similar for the three galaxies.
The sodium lines show strong trends with radius, whereas Mg
and the calcium triplet are nearly constant with radius.
}
\label{radvar.fig}
\end{figure*}

Extracted spectra at $R=0$ and $R\sim R_{\rm e}/2$ are shown
in Figs.\ \ref{full_spectra1.fig} --
\ref{full_spectra3.fig}, for the three primary galaxies.
The $R\sim R_{\rm e}/2$ spectrum is the average of the three radial bins
that are closest to $R=\pm R_{\rm e}/2$.
As we show later, this is the approximate radius where
the IMF is no longer bottom-heavy but consistent with
that of the Milky Way.
The spectra were de-redshifted and smoothed to a common
resolution of 450\,\kms, so they can be compared directly.
The chosen wavelength ranges do not cover the full extent
of the spectra but correspond approximately to the regions that
were used in the stellar population modeling (see Conroy et al.\ 2017). 
There are several obvious differences between the inner
spectra and outer spectra. The Na lines,
and particularly \nad, are stronger in the center than at
$R=R_{\rm e}/2$. For NGC\,1407 and NGC\,1600
the \feh\ band is also more prominent in the central
aperture. 
In this study we do not analyze the strength of individual
absorption features, as their interpretation is not straightforward
(see, e.g., Conroy \& van Dokkum 2012a). Nevertheless, given its
well-known IMF sensitivity we discuss measurements
of FeH in Appendix C.

\subsection{Modeling}

The spectra were fitted with stellar population synthesis models,
following the procedures described in {Conroy} \& {van Dokkum} (2012b).
Only spectra with a median S/N ratio $>30$\,\AA$^{-1}$ in the red
were used.
The S/N is 200--500 in the center (depending
on the galaxy) and falls off to
values near the limit in the outermost bins.
For the three primary galaxies there are typically
20 radial bins
that satisfy this criterion; for the other three galaxies there
are typically 10. In total, 92
spectra were used. 

Almost every aspect of the models has been updated since
the analysis in {Conroy} \& {van Dokkum} (2012b).  The changes are briefly
summarized here, and described in more detail 
in Conroy et al.\ (2017).
The most important difference is that the models cover a
large range of metallicity and age, owing to the use of a greatly
expanded stellar library (Villaume et al.\ 2016). This is important as we
aim to separate abundance gradients from IMF gradients. The
model ages range from 1\,Gyr to 13.5\,Gyr and [Z/H] ranges from
$-1.5$ to $+0.25$. 
Furthermore, the MIST stellar isochrones are used
({Choi} {et~al.} 2016), which cover a wide range of ages, masses,
and metallicities. 
The expanded stellar library uses the
M dwarf library of {Mann} {et~al.} (2015) and newly obtained
near-IR spectroscopy of 283 stars (which already had
optical spectra from the MILES library {S{\'a}nchez-Bl{\'a}zquez} {et~al.} 2006). 
A spectral interpolator is used to provide spectra on
a continuous grid of $T_{\rm eff}$, $\log g$, and metallicity.
The library and interpolator are presented in Villaume et al.\ (2016).
Finally, the elemental response functions have been revised,
using updated atomic and molecular line data. 

The models have 36 free parameters, including 17 individual
elemental abundances and several nuisance parameters.
A list of all parameters is given in Appendix A, and
details are given in Conroy et al.\ (2017).
Two of the nuisance parameters are the weight and
temperature of a hypothetical
hot star component; we verified that the fit results are
nearly identical if this component is turned off.
Among several ``data'' parameters is a multiplicative factor
that is applied to all the errors and a scaling factor that
is applied to a residual telluric absorption spectrum.
Following our previous papers
the IMF has two free parameters, $x_1$ and $x_2$,
which are the logarithmic slopes of the IMF in the
mass ranges $0.08<M/{\rm M}_{\odot}<0.5$ and $0.5<M/{\rm M}_{\odot}
<1.0$
respectively. The IMF is assumed to have
the {Salpeter} (1955) slope of 2.35
at $M>1.0 {\rm M}_{\odot}$. In Conroy et al.\ (2017) we analyze
more complex forms of the IMF.
The models are fit using a Markov chain Monte Carlo
algorithm ({Foreman-Mackey} {et~al.} 2013), after broadening them to the
(wavelength-dependent) instrumental resolution (see \S\,2.3.2).

The best-fitting models are shown in red in Figs.\
\ref{full_spectra1.fig} -- \ref{full_spectra3.fig},
after shifting them to the rest-frame and
smoothing to a resolution of $\sigma=450$\,\kms. Note that
the fits were performed on the original spectra, not on the
smoothed spectra.
The fits
are generally excellent, but there are systematic
differences between the models and the data that exceed the
expected photon noise. This is demonstrated in
Fig.\ \ref{res.fig}, which shows the ratio between the
data and the best-fitting model, for 
all six galaxies and all radial bins. The residuals are
not consistent with random noise but are
highly correlated, with the residuals from each of
the 92 spectra showing the
same rest-frame wavelength dependence. The green
line shows the median as a function of wavelength.
The rms of these systematic residuals at a resolution
of $\sigma = 450$\,\kms\ is 0.50\,\%
from 4000\,\AA\ -- 5000\,\AA, 0.23\,\% from 5000\,\AA\ --
6300\,\AA, 0.22\,\% from 8000\,\AA\ -- 8900\,\AA, and
0.20\,\% from 9650\,\AA\ -- 10070\,\AA.
We also show the residual after subtracting the green line. There
are virtually no features in these ``residuals of the residuals'',
which means that we model the {\em variation} of the spectra (from
galaxy to galaxy and
as a function of
radius) extremely well.

Qualitatively similar behavior was seen in {Conroy} \& {van Dokkum} (2012b)
for individual early-type galaxies and in {Conroy}, {Graves}, \& {van  Dokkum} (2014)
for SDSS stacks. The residuals are probably not due
to problems in the data, as the
six galaxies have different radial velocities: if the
systematic residuals
were related to the sky subtraction, the telluric absorption
correction, or the response function they would line
up in the observed frame, not the rest-frame.
It is also unlikely that they are caused by errors in
the line profiles. Although the assumption of Gaussian
profiles is a simplification (see, e.g., {van der Marel} \& {Franx} 1993),
there is no correlation between the strength of
absorption features and the amplitude of the residuals.

The most likely cause is deficiencies in the stellar population
synthesis models at the $\sim 0.2$\,\% level.  The strongest
residuals are at
$\approx 8200$\,\AA\ and $\approx 8450$\,\AA. These are probably
related to TiO; note that the
feature at $\approx 8200$\,\AA\ is redward of the
\nai\ doublet and coincides with a TiO band head at
$\lambda 8205$\,\AA. The interplay between
this TiO feature and Na\,{\sc i} is demonstrated explicitly
in the inset of Fig.\ 9 in {van Dokkum} \& {Conroy} (2012).
In Appendix B we show that these residuals correlate only weakly with
metallicity, radius, and the IMF mismatch parameter. In particular,
we show that the variation in the residuals is significantly weaker
than the signal from IMF variations.
The residuals in the present study do not correlate very well
with those in {Conroy} \& {van Dokkum} (2012b) or {Conroy} {et~al.} (2014).
For example, the main deficiency in
the red in {Conroy} \& {van Dokkum} (2012b) 
was that the models underpredicted the strength
of the calcium triplet lines, whereas in the present study
the main residuals are at the locations of TiO bandheads.\footnote{Compared
to the 2012 work we also
improved the treatment of instrumental broadening in
the present study. This may be relevant for the residuals near the calcium
triplet.}
This is perhaps not surprising given
the many changes to the
stellar population synthesis models since our previous
papers.

\subsection{Radial Variation in the Spectra}

In this paper we are primarily concerned with the {\em variation}
in the spectra as a function of radius. In Fig.\ \ref{radvar.fig}
we show the ratio of the central spectrum to that at $r\sim
0.5 R_{\rm e}$, that is, the ratio of the two spectra shown for each
galaxy in Figs.\ \ref{full_spectra1.fig} -- \ref{full_spectra3.fig}.
These ratio spectra show which spectral features have strong
radial gradients and which are approximately constant. Expressed
as a ratio, \nad\ shows the largest variation of all individual
spectral features. Its equivalent width in the ratio
spectrum ranges from 0.8\,\AA\ for NGC\,1600 to 2.3\,\AA\ 
for NGC\,2695. Na\,D is a well-known interstellar medium (ISM)
line, but its increase toward small radii
is almost certainly stellar in origin. In HST
images NGC\,1600 has no visible dust absorption
in its central regions ({van Dokkum} \& {Franx} 1995), and neither has
NGC\,1407.\footnote{We visually
inspected Advanced Camera for Surveys
images of NGC\,1407. NGC\,2695 has not been observed with HST.}
Furthermore,
the decrease of
Na\,D with radius is gradual rather than abrupt, and there
is no obvious kinematic difference between this line and
the other absorption lines (see {Schwartz} \& {Martin} 2004).

The \nai\ doublet also varies strongly with radius, as do many
other spectral features, particularly in the blue. 
The \mg\ feature and the calcium triplet
lines do {\em not} show much variation.
It is striking how complex the ratio spectra are, and how
different from the actual spectra.
This illustrates the difficulty of interpreting spectral index
measurements and the power of full spectrum fitting. For
each galaxy, the
red line shows the ratio of the best-fitting models. The
models generally fit the variation in the spectral
features within the (correlated) errors, as shown explicitly
for the FeH band in Appendix C.
Furthermore, they not
only reproduce the changes (and lack of changes) in strong
features such as Mg and NaD, but also the behavior
of the spectra on all scales.

\section{Radial Gradients in Stellar Populations and the IMF}
\label{trends.sec}

\subsection{Stellar Population Parameters}
\label{starpops.sec}

\begin{figure*}[htbp]
  \begin{center}
  \includegraphics[width=0.95\linewidth]{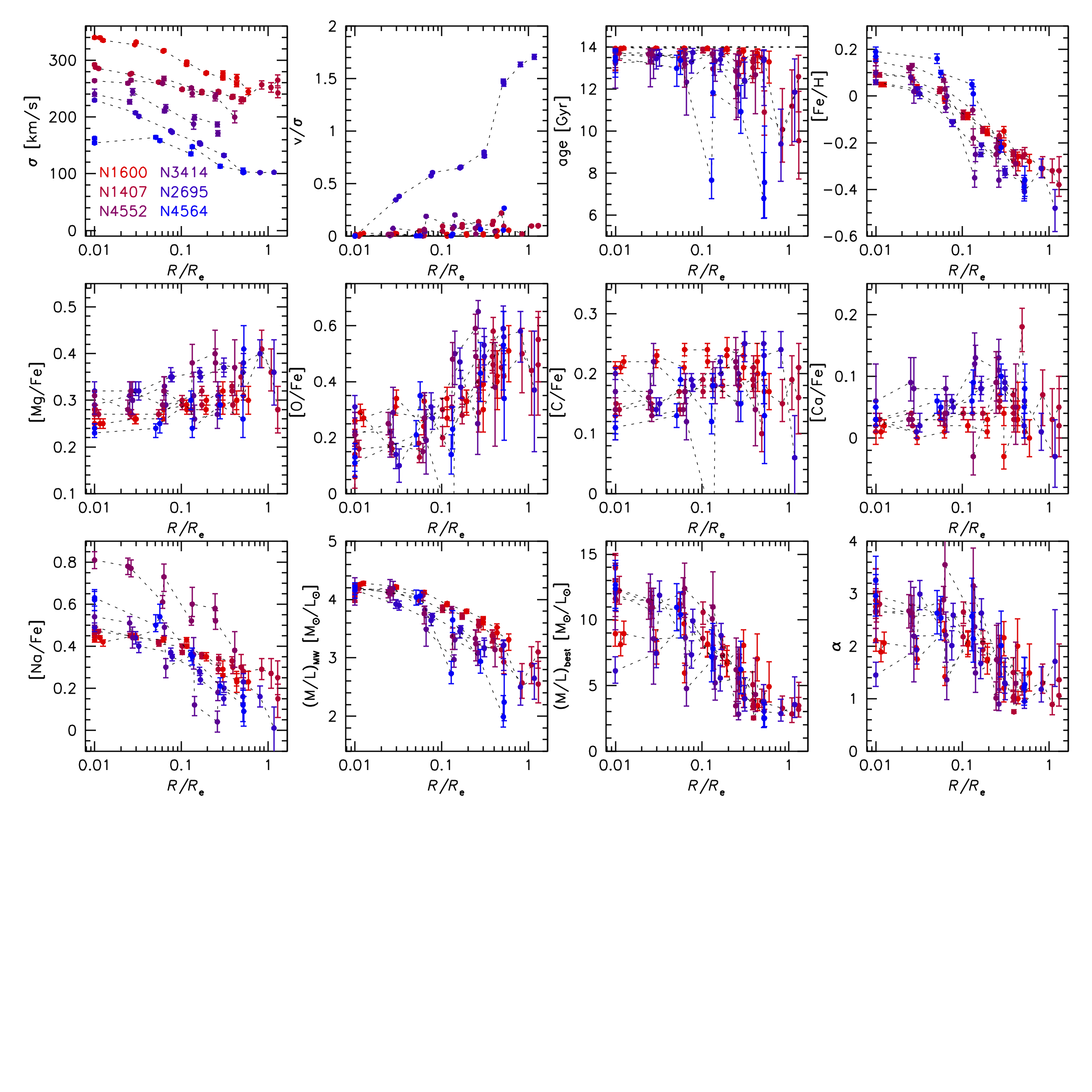}
  \end{center}
\vspace{-5cm}
    \caption{
Stellar population gradients, as derived from full spectrum
fitting. Data points at $R=0$ were placed at $\log R/R_{\rm e}
=-2$. The six galaxies have similar gradients, showing
the same qualitative behavior. 
}
\label{starpops.fig}
\end{figure*}

We now turn to the measured values for the kinematics,
stellar abundance ratios, ages, and mass-to-light ($M/L$)
ratios of the galaxies, as a function of radius.
The model has 36 free parameters; 11 of these
are shown in
Fig.\ \ref{starpops.fig}. From left to right and top to
bottom, the panels show the velocity dispersion (taking
instrumental broadening and the model resolution into
account); the degree of rotational support; the age;
the iron abundance; the abundances of Mg, O, C, Ca, and Na
with respect to iron; the $M/L_r$ ratio for a {Kroupa} (2001)
IMF; the $M/L_r$ ratio for the best-fit IMF; and the
ratio of these $M/L$ ratios $\alpha$.
Errorbars indicate the 16$^{\rm th}$
and 84$^{\rm th}$ percentiles of the posterior probabilities.

\begin{figure*}[htbp]
  \begin{center}
  \includegraphics[width=0.75\linewidth]{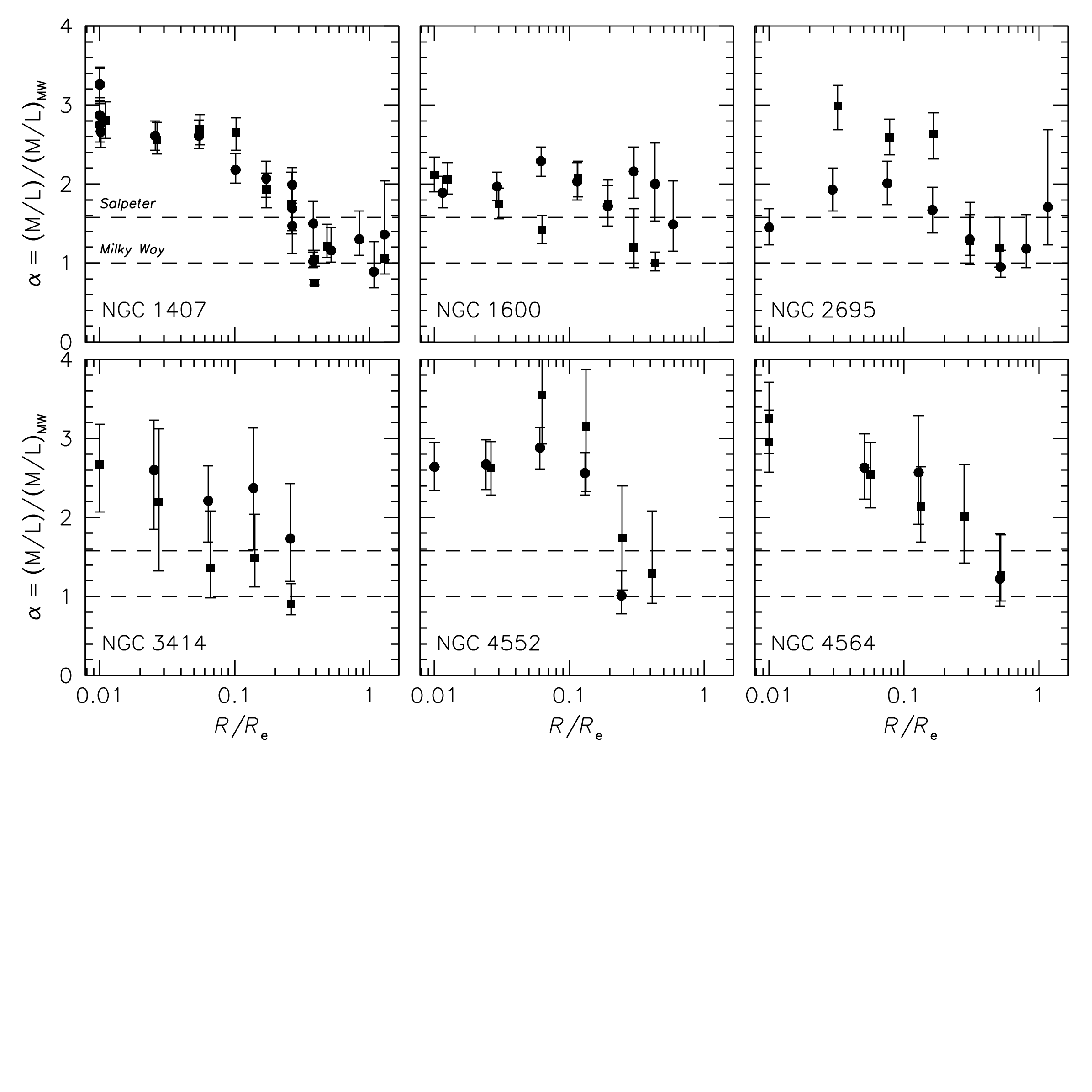}
  \end{center}
\vspace{-4.8cm}
    \caption{
Radial gradients in the IMF parameter $\alpha$, for all six
galaxies. Squares and circles indicate the two different sides of
the galaxies.
All galaxies have a bottom-heavy IMF in their centers,
compared to the IMF of the Milky Way. The IMF becomes less
bottom-heavy with increasing radius, and is consistent with
that of the Milky Way at the last-measured point.
}
\label{imf_ind.fig}
\end{figure*}

\begin{figure*}[htbp]
  \begin{center}
  \includegraphics[width=0.75\linewidth]{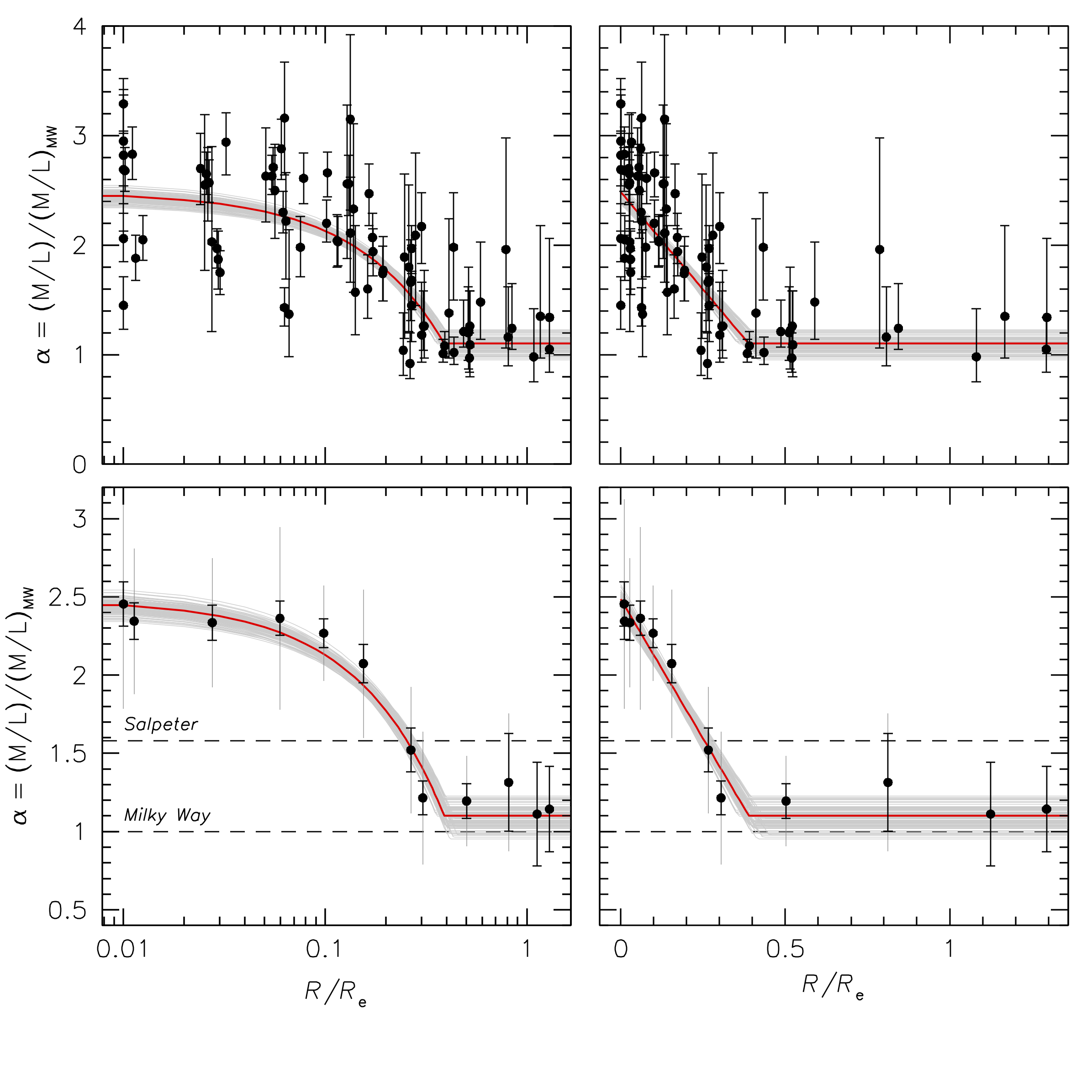}
  \end{center}
\vspace{-1.5cm}
    \caption{
Average radial gradient in the IMF parameter for the six
galaxies, with a logarithmic (left) and linear (right)
$x-$axis. Data points at $R=0$ were placed at $\log
R/R_{\rm e}=-2$. The line has the form
$\alpha = 2.5-3.6(R/R_{\rm e})$ at $R<0.4R_{\rm e}$ and
$\alpha = 1.1$ at $R>0.4R_{\rm e}$. Grey bands indicate 100 random
MCMC samples.
The bottom panels are binned versions of the top panels.
Thin grey errorbars indicate the rms in each bin.
}
\label{alpha_fit.fig}
\end{figure*}

The galaxies span a range of a
factor of two in central dispersion, from
163\,\kms\ to 340\,\kms. Five of the
six galaxies are slow rotators; the exception is
NGC\,2695, which is an S0 galaxy with a prominent disk.
The ages are uniformly high:
all galaxies are older than 10\,Gyr at all measured radii.

A full analysis of the metal line gradients is beyond
the scope of the present paper. Briefly,
the gradients are very similar from one galaxy to the
next, and broadly consistent with previous work
(e.g., {Trager} {et~al.} 2000; {Kuntschner} {et~al.} 2010; {Greene} {et~al.} 2015).
The iron abundance
decreases with radius, from ${\rm [Fe/H]}\approx 0.1$
in the center to ${\rm [Fe/H]}\approx -0.3$ at $R=R_{\rm e}/2$.
The [Mg/Fe] and [O/Fe] ratios increase with radius. The increase
with radius in the $\alpha$-elements is somewhat more pronounced
than previously found for NGC\,1407 ({Spolaor} {et~al.} 2008b), but
we note that these gradients depend sensitively on the assumed relation
between [X/Fe] and [Fe/H] in the stellar population synthesis model.
We use the measured Mg abundances for MILES stars by Milone et al.\
(2011) to derive the [Mg/Fe] vs.\ [Fe/H] relation in our model\footnote{For [O/Fe] we adopt the relation
from Schiavon (2007). For [Ca/Fe] we fit a relation based on
the Bensby et al.\ (2014) data. We assume that Ca, Ti, and Si all
trace one another, and apply the same correction factor to Ti and
Si as we apply to Ca. We assume no corrections are necessary for
the iron-peak elements (Cr, Ni, Cu, V, Mn, and Co) and
that no corrections are necessary for C, N, and Na. This approach
is supported by known trends in the literature; see, e.g.,
Bensby et al.\ (2014) for Na, Cr, and Ni, and the literature
compilation by Kobayashi et al.\ (2006).}; using
the sample of stars of Bensby et al.\ (2014) instead would change the radial
gradient by $\sim 0.1$ dex.
These issues will be discussed further in
a future paper; here we note that the errors do not include the contribution
of these calibration uncertainties.
We also note that the model fits,
and the IMF constraints, are
completely independent of this calibration: the fits measure relative
abundances with respect to the stellar library, and the conversion to
absolute abundances takes place after the fit.

The Ca abundance tracks Fe, as was also seen in previous studies
(e.g., {Saglia} {et~al.} 2002; {Graves} {et~al.} 2007; {Conroy} {et~al.} 2014). The [Na/Fe] ratio
shows a steep negative gradient, again consistent with previous
work ({Boroson} \& {Thompson} 1991, and many other studies).
As [Fe/H] also decreases with radius, the gradient
in the sodium abundance [Na/H] is even steeper; it
decreases from $\sim 0.6$ in the center to $\sim -0.1$ at 
$R\gtrsim R_{\rm e}/2$. This is important
as the \nai\ doublet is a key IMF diagnostic. We will return
to this in \S\,\ref{imftest.sec}.

\subsection{IMF Gradients}

The last three panels of Fig.\ \ref{starpops.fig} show the
key result of this study. The second panel of the
bottom row shows the $M/L$ ratio
as a function of radius when assuming a {Kroupa} (2001)
IMF. The $M/L$ ratio gradually decreases with radius, from
$M/L_r\approx 4$\,M$_{\odot}/$L$_{\odot}$ in the center
to $M/L_r\approx 3$\,M$_{\odot}/$L$_{\odot}$ at $R=R_{\rm e}/2$,
again
consistent with many previous studies (e.g., {Tortora} {et~al.} 2011).
The next panel shows the $M/L$ ratio when allowing the IMF to vary.
For all six galaxies, the central $M/L$ ratio is much higher
for a varying IMF than for a {Kroupa} (2001) IMF. That is,
all galaxies prefer a bottom-heavy IMF that is steeper than
that of the Milky Way at low masses. The final panel shows
the IMF ``mismatch''
parameter $\alpha$, defined as $\alpha\equiv (M/L)/(M/L)_{\rm MW}$.
The IMF parameter falls off steeply, from $\alpha =2-3$ at $R=0$ to
$\alpha \sim 1$ at $r\gtrsim 0.3 R_{\rm e}$.

The $\alpha$ gradients for the individual galaxies are shown
in Fig.\ \ref{imf_ind.fig}. Only measurements with average uncertainties
$\Delta \alpha = (\alpha_{84}-\alpha_{16})/2 <1$  are shown.
The $x-$axis is logarithmic,
to show the individual points more clearly. For the purpose
of this Figure the central aperture ($R=0$) was placed at
$R=0.01 R_{\rm e}$.
At most radii there are two measurements, one for each side of
the galaxy. It is reassuring
that the measurements on each side of the galaxy are
generally consistent within the uncertainties.\footnote{A possible exception
is NGC\,2695, where three adjacent bins show one side to be higher
than the other by 2--3$\sigma$.}
The six galaxies show the same trends:
the IMF is bottom-heavy in the center, and gradually becomes
more bottom-light. The last-measured point is consistent with
the IMF of the Milky Way for all galaxies. 

We parameterize the IMF variation in the following way. We
fit a model of the form $\alpha = a (R/R_{\rm e}) + b$ at
small radii, reaching a plateau of $\alpha = c$ when
$a(R/R_{\rm e})+b \leq c$. Gradients in early-type galaxies
are usually expressed in $\log R$ rather than $R$, but the
advantage of a simple linear function is that it does not
diverge at $R=0$. Using all data points for the
six galaxies we find
\begin{equation}
\label{imf.eq}
\alpha\left(\frac{R}{R_{\rm e}}\right)=\max\
\begin{cases}
2.48^{+0.05}_{-0.05} - 3.6^{+0.3}_{-0.2}
\left(\frac{R}{R_{\rm e}}\right)\\
1.10^{+0.05}_{-0.06}
\end{cases}
.
\end{equation}
This fit is shown in Fig.\ \ref{alpha_fit.fig}.
It is remarkable that the best-fitting value for $c$ is
only 10\,\% larger than the Milky Way IMF; there is
no known aspect of our modeling that prefers the
{Kroupa} (2001)
or {Chabrier} (2003) IMF over other forms.
The rms of the residuals of the fit
is 0.41, a factor of 1.3 higher than
the expected scatter from the
formal errors.  This is probably due
to a combination of systematic errors in the
models and galaxy-to-galaxy variation in the IMF
(see \S\,\ref{imftest.sec}).
The transition to the plateau value of $\alpha=1.1$ occurs
at $R=0.4 R_{\rm e}$.
Based on these six galaxies,
we conclude that bottom-heavy IMFs are a phenomenon that
is unique to the centers of massive galaxies, on physical
scales of $\lesssim 1$\,kpc. 

\begin{figure}[htbp]
  \begin{center}
  \includegraphics[width=1.0\linewidth]{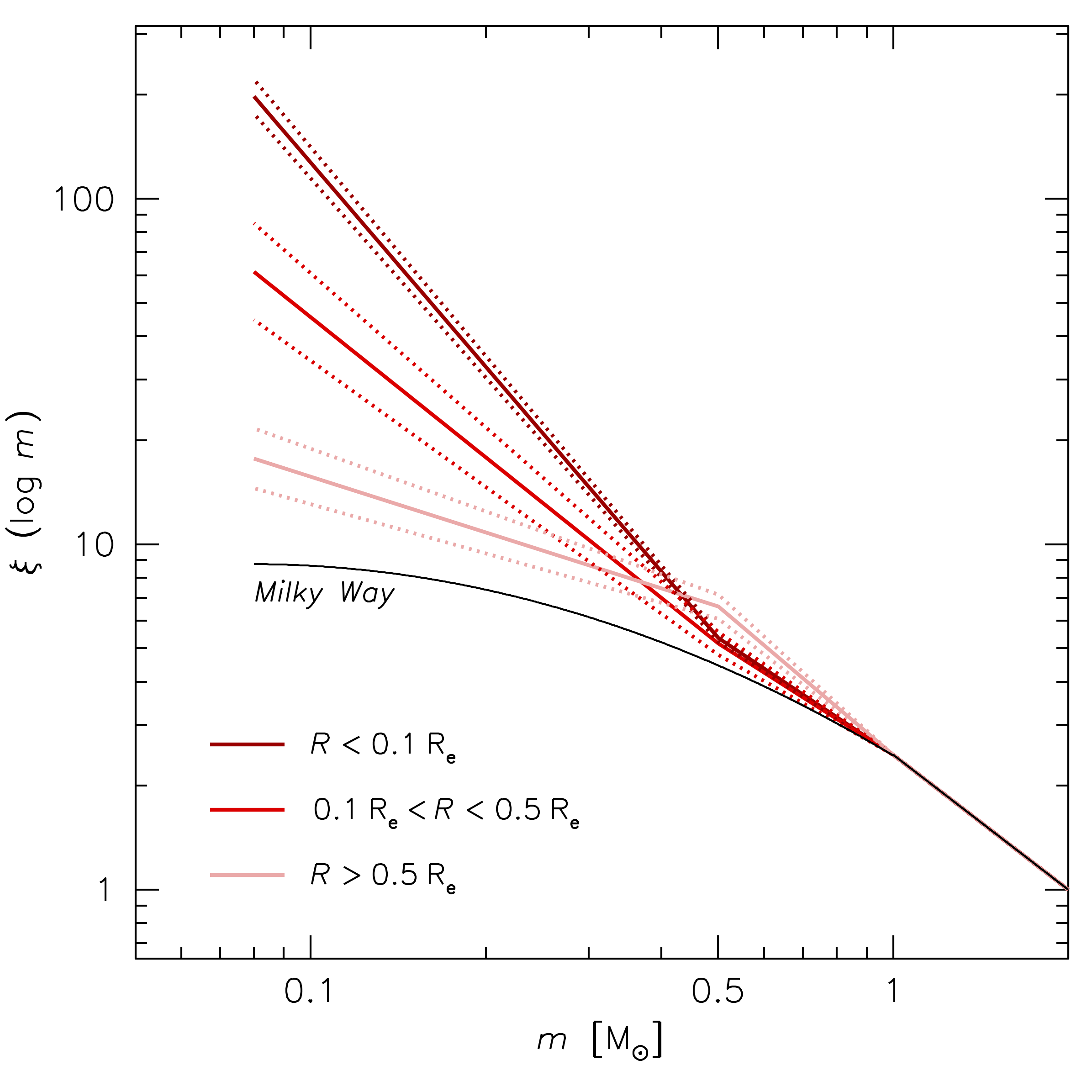}
  \end{center}
\vspace{-0.55cm}
    \caption{
Average form of the IMF in three radial bins,
as determined from the best-fitting logarithmic slopes in
the stellar mass ranges
$0.08$\,\msun\,$<m<0.5$\,\msun\ and $0.5$\,\msun\,$<m<1.0$\,\msun.
The IMFs were normalized so they all have the same number of stars at
$m=1$\,M$_{\odot}$.
The IMF is steeper than the Salpeter form in the center (see
Conroy et al.\ 2017) and similar to the Milky Way IMF at large
radii.
}
\label{imf_shape.fig}
\end{figure}

We express these results in a different way in Fig.\ \ref{imf_shape.fig},
which shows the form of the IMF in three radial bins: $R<0.1R_{\rm e}$,
$0.1R_{\rm e}\leq R\leq 0.5R_{\rm e}$, and $R>0.5R_{\rm e}$. In each
radial bin we determined the averages of the fit parameters $x_1$
and $x_2$, the logarithmic slopes of the IMF in the mass ranges
$0.08$\,\msun\,$<m<0.5$\,\msun\ and $0.5$\,\msun\,$<m<1.0$\,\msun\
respectively (see Appendix A). The average IMF is steep in the center, with
$x_1=2.97\pm 0.05$ and $x_2=2.13\pm 0.04$. At intermediate radii the
form of the IMF is close to Salpeter (which has $x=2.3$ at all masses):
we find $x_1=2.35\pm 0.13$ and $x_2=2.07\pm 0.11$. Beyond $R=0.5R_{\rm e}$
the average IMF has $x_1=1.54 \pm 0.06$ and
$x_2=2.43 \pm 0.12$. This is close to the IMF of the Milky Way:
the Kroupa (2001) form has $x_1=1.3$ and $x_2=2.3$, and
the Chabrier (2003) IMF is shown in Fig.\ \ref{imf_shape.fig}.
An in-depth discussion of the form of the IMF in the center of
NGC\,1407 is given in paper IV in this series (Conroy et al.\ 2017).

\section{Discussion}

\subsection{Disentangling IMF Effects and Abundance Effects}
\label{imftest.sec}

In this paper
we find strong gradients in the IMF of massive early-type galaxies.
These results are qualitatively
consistent with several other recent studies
(e.g., {Mart{\'{\i}}n-Navarro}  {et~al.} 2015a; {La Barbera} {et~al.} 2016).
However,
others have stressed the difficulty of disentangling the effects 
of abundance variations and the effects of the IMF
({McConnell} {et~al.} 2016; 
{Zieleniewski} {et~al.} 2016).
Our modeling allows for independent variations in all the relevant
elements and optimally uses the information content of the full
optical spectra.
Nevertheless, it is a valid question whether the subtle effects of
a changing initial mass function
can really be reliably detected given the (sometimes
dramatic) changes in the stellar abundances.

\begin{figure}[htbp]
  \begin{center}
  \includegraphics[width=0.9\linewidth]{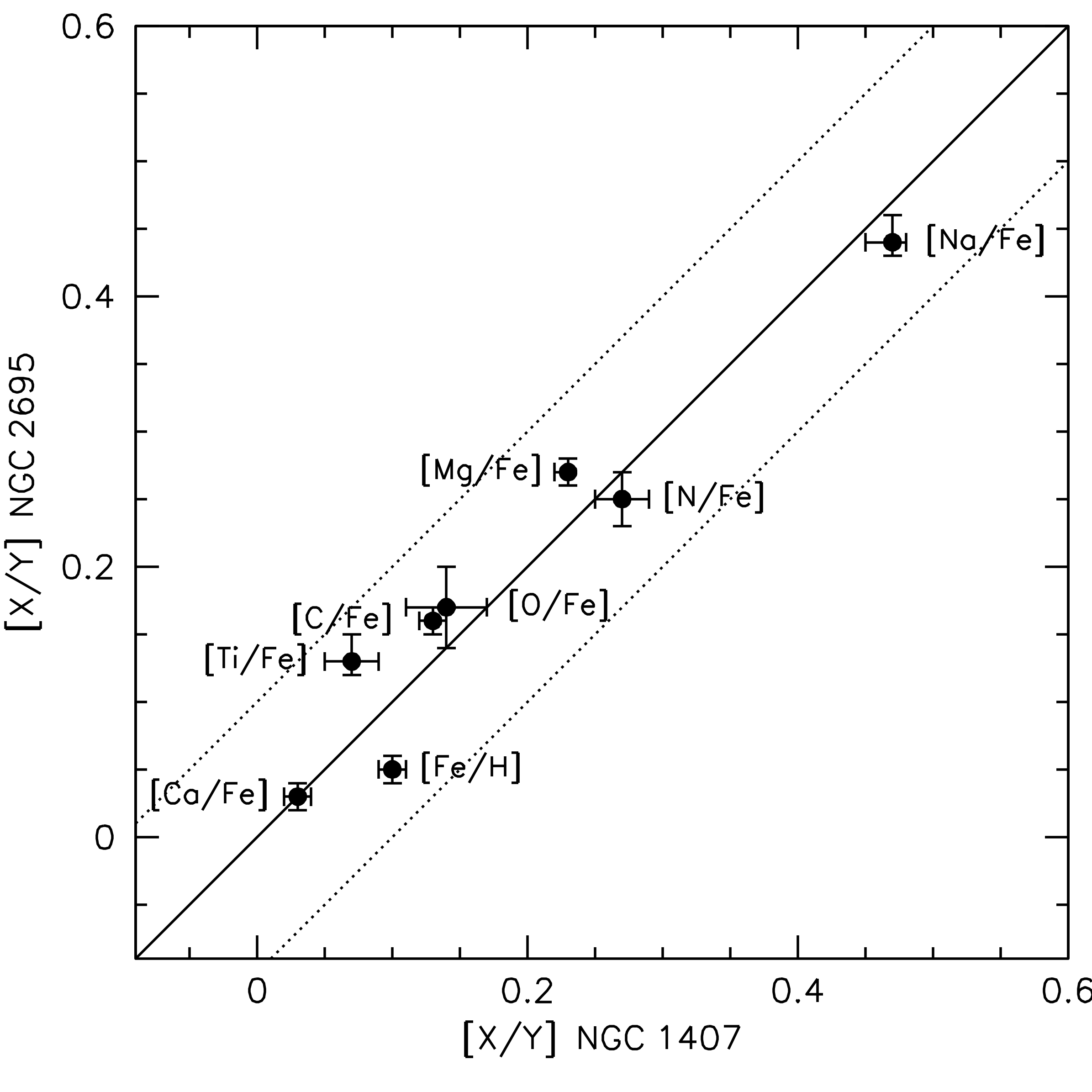}
  \end{center}
\vspace{-0.55cm}
    \caption{
Key stellar abundances of the central spectra of NGC\,1407 and
NGC\,2695. They are very similar, with $\Delta{\rm [X/Y]}<0.1$
for all measured elements. The age difference is also very small
at $\Delta {\rm age}=0.2$\,Gyr. However, the derived IMFs are
quite different: $\alpha=3.3$ for the
NGC\,1407 spectrum and $\alpha=1.9$ for the NGC\,2695 spectrum.
}
\label{index_comp.fig}
\end{figure}

The high S/N spectra presented in this paper
offer the opportunity for a ``semi-empirical'' demonstration
how IMF effects and abundance effects can be distinguished.
We compared the derived stellar population parameters for all 92
spectra and looked for pairs of spectra where the
abundance pattern and age are a close match but the IMF is
very different. This analysis is similar in spirit to the
comparison between massive 
elliptical galaxies in Virgo and M31 globular
clusters that we did in 2011 ({van Dokkum} \& {Conroy} 2011).
Using the criteria
$\Delta {\rm age}<2$\,Gyr, $\Delta {\rm [Fe/H]}<0.1$,
$\Delta {\rm [Na/Fe]}<0.1$, $\Delta {\rm [Ca/Fe]}<0.1$,
$\Delta {\rm [O/Fe]}<0.1$, $\Delta {\rm [Ti/Fe]}<0.1$,
and $\Delta \alpha>1$ we found a single match: the
central 3-pixel aperture of NGC\,1407 and a slightly larger central
aperture (of 5 pixels) of NGC\,2695.
The age is 13.3\,Gyr for the NGC\,1407 spectrum
and 13.7\,Gyr for the NGC\,2695 spectrum. The abundance ratios
of eight elements are shown in Fig.\ \ref{index_comp.fig}.
They are very similar, although it should be noted that
the differences in [X/Fe] are somewhat smaller than the
differences in [X/H], given the 0.05\,dex offset in [Fe/H].
The best-fitting IMF  parameters are (by selection)
quite different, with
$\alpha=3.29^{+0.23}_{-0.25}$ for NGC\,1407 and
$\alpha=1.93^{+0.22}_{-0.19}$ for NGC\,2695.

\begin{figure*}[htbp]
  \begin{center}
  \includegraphics[width=0.85\linewidth]{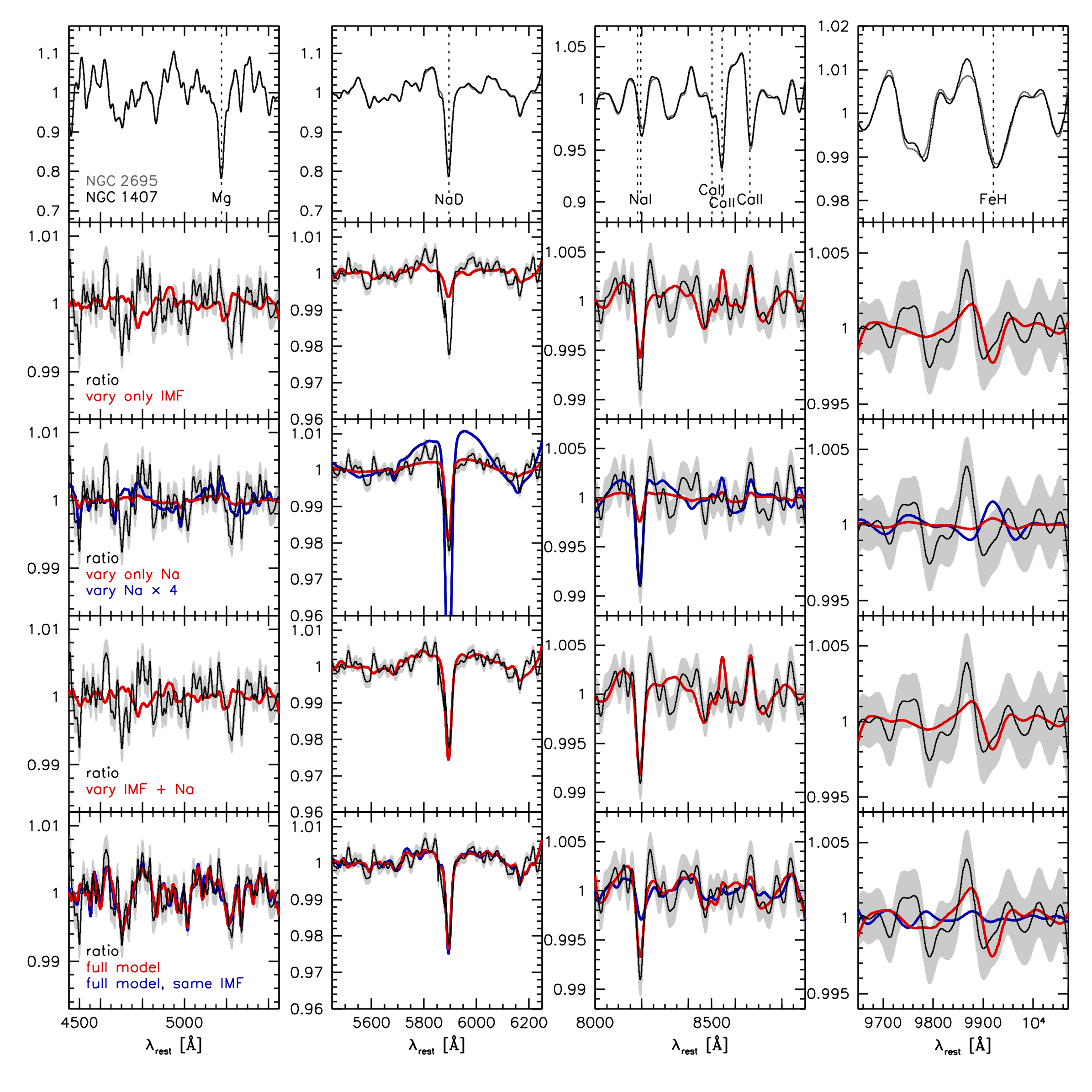}
  \end{center}
\vspace{-0.5cm}
    \caption{
Top row: central spectra of NGC\,1407 and NGC\,2695, 
smoothed to the same resolution. The galaxies have nearly 
identical ages and abundances, but NGC\,1407 has a more bottom-heavy
IMF than NGC\,2695. The spectra are nearly identical.
All other panels show
the ratio between these two spectra, and various model permutations.
Second row: model that only has IMF variation. The fit
to Na\,{\sc i}
is reasonable, but Na\,D is underpredicted. Third row:
model with only Na variation. This model fits Na\,D well but
underpredicts Na\,{\sc i}. The blue model fits Na\,{\sc i}
but overpredicts Na\,D by a large factor. Fourth row:
a combination of Na variation and IMF variation is needed
to reproduce both Na\,{\sc i} and Na\,D. Bottom row: full
model, which
also includes the small differences in all other elements.
The blue model is the full model but with the IMF held fixed to
the average IMF of the two galaxies. Only the full model
with a varying IMF (red) fits the data.
}
\label{matched_specs.fig}
\end{figure*}

The two spectra are shown in the top panels of Fig.\
\ref{matched_specs.fig} (black and grey).
They are, as expected, very similar. The best-fitting models
are overplotted in red and blue. Rows 2, 3, 4 and 5 of
Fig.\ \ref{matched_specs.fig} all show the ratio of the two spectra
in black, with the errors in grey. The ratio spectrum is flat to
$\approx 0.5$\,\% over most of the wavelength range.
The two most prominent exceptions are the \nad\ doublet and
the \nai\ doublet, which are both stronger in NGC\,1407 than
in NGC\,2695. There are also systematic differences at
the 0.5\,\% level at the location of
the $\lambda 8665$\,\AA\ Ca triplet line, near the \feh\
band, and in the blue near 4600\,\AA. 

We created ad hoc models for NGC\,1407 and NGC\,2695
to investigate whether we can isolate
the IMF features in the ratio spectrum. First, we made two
models that differ only in their IMF.
That is, the model for NGC\,1407 has $\alpha = 3.3$ and
the model for NGC\,2695 has $\alpha=1.9$ but in all other
respects they are identical, with all model parameters
except the IMF set to the averages of the two galaxies.
The red line in the second row of Fig.\ \ref{matched_specs.fig}
shows the ratio of these two models. These IMF-only models do
well in the red, fitting the Na\,{\sc i} doublet,
the Ca triplet region, and the FeH band reasonably well.
The residuals in the blue are not well matched by this model;
this is expected as the blue spectrum is not very sensitive
to variation in the contribution from low mass stars.
More importantly,
IMF-only models fail to account for the large difference
in Na\,D between the two galaxies:
only $\sim 1/3$ of its depth in the ratio spectrum
can be accounted for by the IMF.

Although the spectra were selected to have {\em similar} abundance
ratios, they are of course not {\em identical}.
Of particular
relevance is that the sodium abundance
of NGC\,1407 is slightly higher than that of NGC\,2695.
NGC\,1407 has ${\rm [Na/Fe]}=0.47$ and NGC\,2695 has
${\rm [Na/Fe]}=0.44$. Taking the [Fe/H]
ratios into account,
we have ${\rm [Na/H]}=0.57$ for NGC\,1407 and
${\rm [Na/H]}=0.49$ for NGC\,2695, a difference of 0.07\,dex.
In the third row
of panels of Fig.\ \ref{matched_specs.fig} we show the ratio
of two models that only differ in their sodium abundance.
The model for NGC\,1407 has ${\rm [Na/H]}=0.57$ and 
the model for NGC\,2695 has ${\rm [Na/H]}=0.49$, and all
other model
parameters, {\em including the IMF}, are identical
and set to the average of the
best-fitting models of the two galaxies. The red line
fits the Na\,D doublet well, but underpredicts the dwarf-sensitive
Na\,{\sc i} doublet by a factor of $\sim 3$. This sodium-only
model also fails to fit the FeH region of the ratio spectrum.

The blue line in these panels reflects an attempt to fit the
Na\,{\sc i} doublet by artificially increasing the Na abundance.
The response is approximately linear, and we fit a model
ratio spectrum of the form $S'=1+f(S-1)$, with $S$ the
ratio spectrum for the actual Na abundances of the two
galaxies and $f$ a free parameter. The blue line has
$f=4$, and it provides a good fit to \nai, by construction.
However, the fit to \nad\ is catastrophic; unsurprisingly,
given that the red line provided a good fit to this feature,
it overpredicts Na\,D by a factor of $f$.
We conclude that the
difference in the observed strength of the Na\,{\sc i} line
between the two galaxies
cannot be attributed to a difference in sodium abundance.
Spectral coverage of the Na\,D doublet is critical:
both Na\,D and Na\,{\sc i}
are senstive to the IMF and the Na abundance, but
Na\,D is {\em mainly}
sensitive to the Na abundance and Na\,{\sc i} is
{\em mainly} sensitive to the IMF (as shown in the second and
third row of Fig.\ \ref{matched_specs.fig}). We note that any ISM
contribution to Na\,D would imply that the IMF is even
more bottom-heavy than what we infer in this paper: it
would mean that the Na abundance is lower, which in turn
implies that more low mass stars are needed to reproduce
the observed strength of \nai.

The models in the fourth row of panels show explicitly
that the combination of a varying IMF and a small difference
in the Na abundance can reproduce both Na\,{\sc i} and Na\,D,
as well as the FeH region. We now created a model for
each galaxy that has
the best-fit IMF and the best-fit Na abundance for that
galaxy, with all other model parameters set to the averages
for the two galaxies. The ratio of these models fits both
Na lines, as well as the overall shape of the spectrum in the
red.

In the bottom row of
panels we show the ratio of the full
models, with the parameters for each galaxy
set to the best fitting values for that galaxy. The
model ratio spectrum is an excellent fit throughout
the spectral range. It is now clear that the residuals
in the blue are not noise but due to the small differences
between the galaxies
in [Fe/H], [Ti/Fe], and the other model parameters.
The rms residuals are 0.19\,\% from 4500\,\AA\ -- 6200\,\AA,
0.13\,\% from 8000\,\AA\ -- 8900\,\AA, and 0.11\,\%
from 9650\,\AA\ -- 10070\,\AA.
As a final test we allow all parameters to vary for both galaxies
except the IMF, which we fix to the best-fitting value for NGC\,2695.
We then take the ratio of this constrained full model fit to NGC\,1407
to the full model fit for NGC\,2695.
The result is the blue line in the bottom panels. The model ratio
spectrum fits the blue reasonably well, but underpredicts Na\,{\sc i}
by a factor $\sim 3$. It also fails to fit the FeH region, and is a
poorer fit to the CaT lines than the red line.
We conclude that our models cannot fit the ratio spectrum unless the
IMF is allowed to vary.

\subsection{What Parameter Best Predicts the Local IMF?}

Throughout this paper we have analyzed the IMF as a function
of $|R/R_{\rm e}|$, the absolute
distance from the center of the galaxy in
units of the half-light radius. This is the
most straightforward choice given that we extract spectra as a function
of radius and the galaxies all have different sizes. However, as
we have not only measured the local IMF but also the local velocity
dispersion, age, and elemental abundances we can ask whether any of
these parameters correlates better with the IMF mismatch parameter
than $R/R_{\rm e}$ does (see also, e.g., {La Barbera} {et~al.} 2016).

\begin{figure}[htbp]
  \begin{center}
  \includegraphics[width=0.95\linewidth]{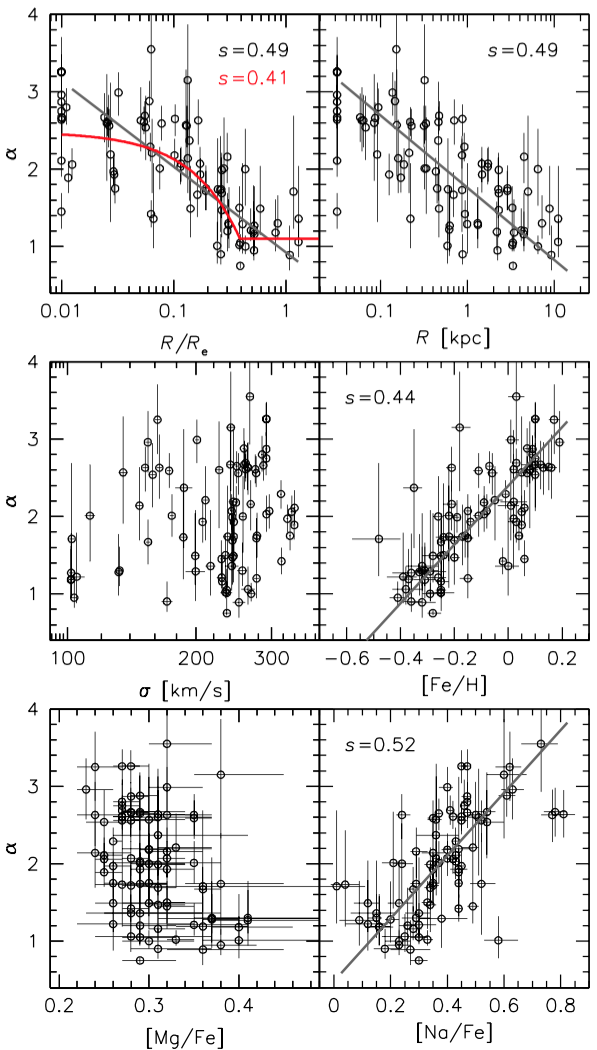}
  \end{center}
\vspace{-0.3cm}
    \caption{
IMF mismatch parameter $\alpha$ as a function of $R/R_{\rm e}$,
radius in kpc, velocity dispersion, [Mg/Fe], [Fe/H], and [Na/H].
The red line is Eq.\ \ref{imf.eq}. Grey lines are simple powerlaw
fits, for all parameters that have a correlation coefficient
$|r|>0.5$. The scatter in the residuals from these fits is indicated
in each panel; the uncertainty in the scatter is $\approx 0.04$.
The metallicity predicts
the local IMF as well as the radius does.
We do not find strong correlations between
$\alpha$ and the local velocity dispersion or [Mg/Fe] ratio.
}
\label{control.fig}
\end{figure}

In Fig.\ \ref{control.fig} we show the IMF mismatch parameter $\alpha$
as a function of six different parameters. The first is $R/R_{\rm e}$,
for reference. The red curve is Eq.\ \ref{imf.eq}; the residuals from
this fit have an rms scatter of $s=0.41$. The grey line is
a fit of the form $\alpha = a \log (R/R_{\rm e})+b$. This fit has
a higher scatter of 0.49, but it enables us to compare the predictive
power of $R/R_{\rm e}$ to the other parameters. The uncertainty
in $s$ is $\approx 0.04$ for all panels,
as derived from the formal error bars combined with
the sample size. The top right panel shows the relation between
$\alpha$ and physical radius.
The scatter is identical to the relation between $\alpha$ and
$R/R_{\rm e}$.

The second row of Fig.\ \ref{control.fig} shows correlations
with velocity dispersion and [Fe/H].
The central velocity dispersion has often been found to correlate with
the central IMF; examples are
{Treu} {et~al.} (2010), {Conroy} \& {van Dokkum} (2012b),
{Cappellari} {et~al.} (2013), and {La Barbera} {et~al.} (2015).
Here we find, somewhat
surprisingly, that the local dispersion is not a 
predictor of the local IMF: the correlation coefficient is only $r=0.23$.
The correlation with [Fe/H], on the other hand,
is tight, with a correlation coefficient of
0.78 and a scatter of 0.44.

In the bottom panels we show relations with [Mg/Fe] and [Na/Fe].
In {Conroy} \& {van Dokkum} (2012b) it was
found that the [Mg/Fe] ratio correlates 
with $\alpha$, and
we suggested that the star formation time scale might be a key
driver of the form of the IMF. 
Here we do not find a strong correlation between $\alpha$ and
[Mg/Fe]. There is, in fact, a weak anti-correlation,
reflecting the mild increase of
[Mg/Fe] with radius in Fig.\ \ref{starpops.fig}.\footnote{Note
that this does not rule out
the existence of a correlation between the central [Mg/Fe] and the
central value of $\alpha$.}
We note that the existence of a strong
positive correlation between $\alpha$ and
[Mg/Fe] had  been called into question by
{Smith} (2014) and by {La Barbera} {et~al.} (2015).
The last parameter that we consider is [Na/Fe],
which shows a positive relation with $\alpha$ but with significant
scatter ($s=0.52$). All relations between $\alpha$ and
[X/Fe] show larger scatter than the 
relations between $\alpha$ and [X/H]. This suggests that the
overall metallicity, rather than a specific element, is
coupled with the form of
the IMF (see also {Mart{\'{\i}}n-Navarro}  {et~al.} 2015c).
However, we cannot determine this conclusively from our data:
the correlation coefficient and scatter
for the relation between $\alpha$ and [Z/H] are
not significantly different from the relation between
$\alpha$ and [Fe/H].

Finally, 
we caution that these results may partially be driven by scatter
between the six galaxies in our small sample; as an example,
the variation in velocity dispersion between the galaxies is larger
than the typical variation within each individual galaxy. Larger
samples are needed to explore these issues further.

\subsection{The Luminosity-Weighted IMF in an Aperture}
\label{imf_int.sec}

\begin{figure}[htbp]
  \begin{center}
  \includegraphics[width=0.95\linewidth]{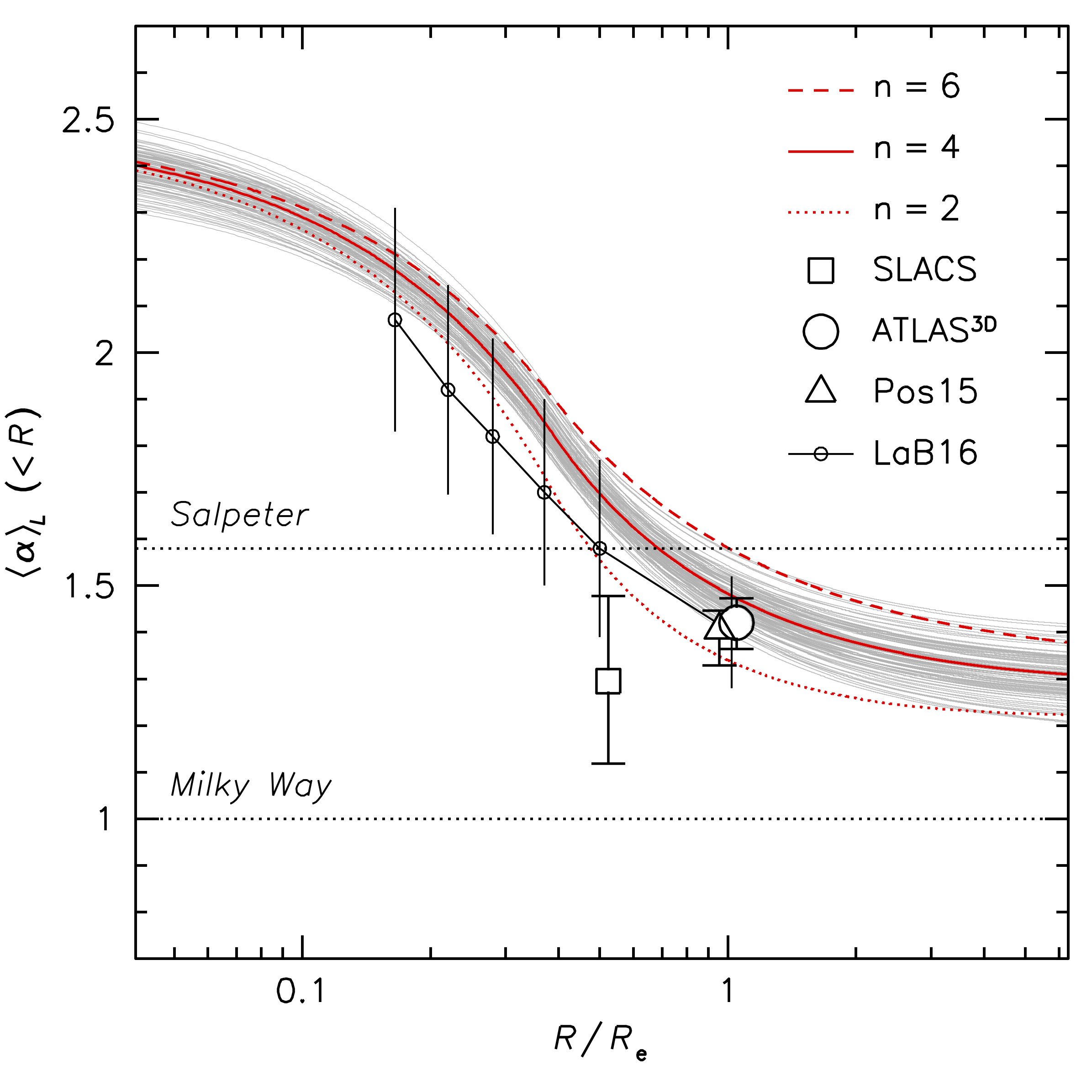}
  \end{center}
\vspace{-0.55cm}
    \caption{
Luminosity-weighted
average IMF parameter $\langle \alpha \rangle_L$ within
an aperture of radius $R$, calculated from Eq.\ \ref{imf.eq}.
The solid red curve
is for a {Sersic} (1968) index $n=4$ and the broken curves
are for $n=2$ (dotted) and $n=6$ (dashed). 
Large open symbols are determined from published
relations between $\langle
\alpha \rangle$ and $\sigma$, using the average dispersion of
the six galaxies in our sample.
The square is derived from
strong gravitational lenses ({Treu} {et~al.} 2010). The circle is
from dynamical modeling of nearby early-type galaxies
({Cappellari} {et~al.} 2013). The triangle is from
{Posacki} {et~al.} (2015), who combined ATLAS$^{\rm 3D}$ and
SLACS data. 
Small open symbols are for a single massive galaxy,
and are taken from
{La Barbera} {et~al.} (2016). Overlapping points are offset slightly in $R$,
for clarity.
All results are in reasonable agreement, given the differences
in methods.
}
\label{imf_int.fig}
\end{figure}

An implication of the existence of IMF gradients is that the
measured IMF depends on the aperture that is used in the
analysis. As noted in \S\,1 this complicates comparisons
between different techniques, such as gravitational lensing
and stellar population synthesis modeling. Using the
form of the gradient given in Eq.\ \ref{imf.eq}
we calculate
the average luminosity-weighted IMF within apertures of
different radii. The galaxies span a range in Sersic index
({Spolaor} {et~al.} 2008a; {Krajnovi{\'c}} {et~al.} 2013), and we calculate the average
IMF for $n=2$, $n=4$, and $n=6$.
The results are shown in Fig.\
\ref{imf_int.fig}.

The average luminosity-weighted IMF parameter is $>1.6$
(i.e., heavier than the Salpeter form) only for
apertures $R\lesssim 0.6R_{\rm e}$. Within 
$R<R_{\rm e}$
we find $\langle \alpha \rangle_L = 1.3-1.5$, depending
on the form of the surface brightness profile. 
Our results
are consistent with those of {La Barbera} {et~al.} (2016),
who analyzed a single massive galaxy in a similar way
(small circles in Fig.\ \ref{imf_int.fig}).
We conclude
that the IMF in apertures that contain most of the light
is lighter than the Salpeter form, and only mildly heavier
than 
a {Kroupa} (2001) or {Chabrier} (2003) IMF.

This aperture-dependence may explain some of the discrepancies in IMF
studies of massive galaxies
in the literature. In particular, dynamical and lensing studies
typically find values for $\alpha$ that are in between the {Kroupa} (2001)
and the {Salpeter} (1955) forms ($1.0 \lesssim \alpha \lesssim 1.6$),
whereas stellar population studies
in much smaller apertures have found $\alpha \sim 2$ and even higher
(e.g., Conroy \& van Dokkum 2012b; Conroy et al.\ 2017).
We can test whether our results are consistent with
dynamical and lensing studies. First, we determine the luminosity-weighted
velocity dispersion as a function of aperture radius.
For each
galaxy the observed relation between $\log \sigma$ and $\log (R/R_{\rm e})$ is
fitted
with a powerlaw and extrapolated to $R=R_{\rm e}$. The luminosity-weighted
dispersion is then obtained by integrating these fits, weighted by
an $r^{1/4}$ law. The mean luminosity-weighted dispersion of the six
galaxies is  $215\pm 23$\,\kms\ within $R_{\rm e}$ (and very similar
within $0.5R_{\rm e}$).
Next, we use published relations between $\langle \sigma \rangle_L$
and $\langle \alpha\rangle_L$ to compare our results to lensing and
dynamical studies. The open square in Fig.\ \ref{imf_int.fig} is
for the SLACS sample of strong lenses in {Treu} {et~al.} (2010); the Einstein
radius of these objects is $R_{\rm E} \approx 0.5R_{\rm e}$. The
circle is from a dynamical analysis of the ATLAS$^{\rm 3D}$ sample
of nearby early-type galaxies ({Cappellari} {et~al.} 2013). The triangle
is from a recent joint analysis of the SLACS and ATLAS$^{\rm 3D}$
samples ({Posacki} {et~al.} 2015). The errorbars reflect only the
formal uncertainty
in the average luminosity-weighted velocity dispersion, and neglect all
other random and systematic errors.
Nevertheless,
the dynamical and lensing results are in excellent
agreement with the gradients
that we derive in this paper.

Four galaxies in our sample are part of the ATLAS$^{\rm 3D}$ survey, enabling
a direct comparison. For the galaxies NGC\,2695, NGC\,3414, NGC\,4552,
and NGC\,4564 {Cappellari} {et~al.} (2013) find $\langle
\alpha \rangle_L \approx 1.4$ with very little scatter. This is, again,
in very good
agreement with the red curve in Fig.\ \ref{imf_int.fig}. If,
instead,
we had compared the ATLAS$^{\rm 3D}$ values to our results
for the centers of the galaxies,
we would have concluded that the studies disagree with
one another (see also {Lyubenova} {et~al.} 2016).

\section{Summary and Conclusions}

Using a new suite of stellar population synthesis models
and high quality data for six galaxies we find that the
stellar IMF in massive galaxies is a strong function of
radius. The IMF is bottom-heavy in the centers
of the galaxies and reaches the approximate Milky Way form
at $R\gtrsim 0.4R_{\rm e}$. This result is consistent with
several recent studies that used different models and were based
on index measurements rather than on full spectrum fitting
(see {Mart{\'{\i}}n-Navarro}  {et~al.} 2015a; {La Barbera} {et~al.} 2016).
It is also consistent with earlier work on the centers
of the most massive galaxies ({van Dokkum} \& {Conroy} 2010; {Conroy} \& {van Dokkum} 2012b).
Given all the changes in our models and the vast improvements
in data quality it is remarkable that we reached similar conclusions
in 2010 as we do today. The reason
is probably that we began by targeting galaxies with very
large velocity dispersions and (as it turned out) very bottom-heavy
IMFs, which are least sensitive to the details of the modeling.
The agreement is not universal, however.
{McConnell} {et~al.} (2016) use very similar data, but
conclude that the observed radial trends
in two early-type galaxies can be explained entirely by abundance
gradients. It may be that those two galaxies happen to have Milky
Way IMFs throughout, but that is not a very satisfying explanation.
Modeling of the {McConnell} {et~al.} (2016) data using similar techniques as
employed here may shed more light on this.

Abundance gradients are certainly a concern in all these
studies, as they can mimic IMF effects.
It is well known that the \nai\ doublet is sensitive to
the Na abundance, and as discussed by
Conroy \& van Dokkum (2012a) and {La Barbera} {et~al.} (2016)
the \feh\ band depends on [Fe/H] and, in certain regimes, on
age and [$\alpha$/Fe].
The upshot is that
the two strongest gravity-sensitive features in the
optical window are both difficult, or even impossible, to
interpret in isolation. Allowing non-solar abundance ratios is
critical, particularly for Na. Furthermore, only full spectrum
fitting, or a carefully chosen
combination of different indices (see {Conroy} \& {van Dokkum} 2012a),
can isolate IMF effects.

As the modeling gets more complex it becomes more difficult to
identify the key data that produce a particular result; the
high degree of sophistication in this field carries the
risk that the modeling begins to resemble a black box.
Some of these issues are addressed in Conroy et al.\ (2017), where
we use the exquisite NGC\,1407 data to analyze what specific stellar
mass ranges are constrained by the data.
In the somewhat stand-alone \S\,5.1 we address this
issue in a different way, analyzing the
ratio of two spectra with similar elemental
abundances but a different IMF. This ``semi-empirical''
approach follows earlier work 
on M31 globular clusters and massive ellipticals
({van Dokkum} \& {Conroy} 2011), although in the present case we do not
have an independent constraint on the IMF that serves
as a ``hard'' limit on the results.
In the case of NGC\,1407 and NGC\,2695 we can point to
specific features that drive the outcome, and the analysis
also provides
understanding of what spectral features constrain the IMF in the
rest of the sample. It is fortuitous
that we have two such spectra in our small sample; future
studies could specifically single out such matched pairs for
deep follow-up spectroscopy to increase the sample.

In \S\,\ref{imf_int.sec} we show that the luminosity weighted
IMF within an aperture depends strongly on the aperture size.
Lensing and dynamical studies typically have an effective aperture
of $0.5-1 R_{\rm e}$, and we demonstrate
in Fig.\ \ref{imf_int.fig} that the gradients
we measure are entirely consistent with the SLACS and ATLAS$^{\rm 3D}$
constraints for galaxies with similar velocity dispersions.
These results are consistent with the recent study of {Lyubenova} {et~al.} (2016),
who found that there is good agreement between dynamical and stellar
population measurements of $\alpha$ when measured
in a self-consistent way and in the same aperture.
It should be noted that the absolute $M/L$ determination remains
somewhat model-dependent in the stellar population synthesis work
(see Fig.\ 12 in Conroy \& van Dokkum 2012b). In principle,
the combination
of lensing and dynamics with stellar population synthesis work can
provide a complete description of both the amount of
mass in galaxies and the (stellar and
non-stellar) sources of that mass (see, e.g., Newman et al.\ 2017).

We emphasize here that our analysis focuses on
massive 
galaxies, with $\sigma(<R_{\rm e})\sim 220$\,\kms.
As even the central IMF is Milky Way-like
for low mass galaxies ({Conroy} \& {van Dokkum} 2012b),
they probably have weaker gradients than the galaxies studied here
(see {Mart{\'{\i}}n-Navarro}  {et~al.} (2015a)
and Spiniello et al.\ (2015b) for studies
of IMF gradients in two such galaxies).
The fact that the galaxy-integrated IMF (to $R=\infty$) has $\alpha \approx
1.3$
for our massive galaxy sample may
mean that the galaxy-wide
IMF is not very different from that of the Milky Way for nearly all
galaxies in the Universe. Possible exceptions are the most
compact massive galaxies ({Mart{\'{\i}}n-Navarro}  {et~al.} 2015b; {Conroy} {et~al.} 2013),
and ultra-faint dwarf galaxies which may have
a bottom-light IMF ({Geha} {et~al.} 2013).


The strong gradients that we find have implications for 
various areas of astrophysics, and we touch on a few here.
First, the large gradient in the $M/L$ ratio (see panel $k$
of Fig.\ \ref{starpops.fig}) needs to be taken into account
when measuring the masses of supermassive black holes.
As an example, {Thomas} {et~al.} (2016) use
stellar kinematics to infer the existence of
a black hole in NGC\,1600 with a mass of $1.7\times
10^{10}$\,\msun, among the most massive black holes ever
found. The sphere of influence of this black hole,
defined as the radius within which the black hole mass
equals the stellar mass, is
$\approx 1$\,kpc, or $\approx 0.07 R_{\rm e}$. Within this
radius we find $\alpha \approx 2$ for NGC\,1600, and our
observed IMF gradient would likely lower the derived black
hole mass (see also Extended Data Fig.\ 5 in Thomas et al.\ 2016).
More detailed modeling of the central kinematics of massive
early-type galaxies, using
the stellar $M/L$ constraints derived here,
is required to understand the full effect on derived black hole masses
(see also L{\"a}sker et al.\ 2013).

Another implication is that the stars that are in the centers
of massive galaxies today formed in a very different way than
the stars that are at large radii. As discussed in \S\,1
there is fairly good evidence that the 
central $R=1-2$\,kpc are indeed unique environments, in that they
assembled in a short period of intense star formation at high
redshift. The outer regions were likely accreted at later times,
and could therefore have abundance ratios and IMFs
that more closely resemble low mass satellite galaxies than the
galaxy centers. Turning this around, the presence of IMF gradients could
be viewed as evidence for the kind of two-phase formation models
that were proposed by {Oser} {et~al.} (2010) and others.
In the context of these models, and studies such as {Fang} {et~al.} (2013),
{van Dokkum} {et~al.} (2014), and {Mart{\'{\i}}n-Navarro}  {et~al.} (2015b),
one might expect that the IMF correlates better
with the physical radius than with $R/R_{\rm e}$. Both parameters
correlate equally well in
our small sample (see \S\,5.2).

Our results do strongly suggest that the IMF in compact, massive
galaxies at $z\gtrsim 2$ --- which are thought to be the progenitors
of the cores of today's massive galaxies --- should be heavier than
that of the Milky Way. Although galaxies obviously evolve  over this
timeframe, both mixing due to mergers ({Sonnenfeld}, {Nipoti}, \&  {Treu} 2016) and
projection effects are expected to decrease, not increase, the
observed central
value of $\alpha$ over time.
There may be some tension with observations: the
dynamics of both
star forming and quiescent massive compact galaxies seem to point
to relatively low $M/L$ ratios ({Belli}, {Newman}, \& {Ellis} 2013; {van de Sande} {et~al.} 2013; {van Dokkum} {et~al.} 2015).
The uncertainties are currently too large to place strong constraints,
but this should change with improved measurements of
high redshift galaxies in the JWST era. It may turn out that
the high redshift
galaxies have gradients too, and/or
that we are misinterpreting their galaxy-integrated
kinematics (as may be indicated by recent observations;
Newman et al.\ 2015; Belli et al.\ 2016).  It will also be important
to have better calibrated stellar
$M/L$ ratios at young ages (see {Dutton} {et~al.} 2012).

Finally, our results have implications for star formation theory.
Probably the most important result in this context is the
finding that the IMF can be even heavier than the {Salpeter} (1955)
form, something that had been suggested in recent studies
(e.g., {Chabrier}, {Hennebelle}, \&  {Charlot} 2014). These implications are discussed in
paper IV in this series (Conroy et al.\ 2017).


\begin{acknowledgements}
We thank Nacho Mart{\'{\i}}n-Navarro for comments on the manuscript,
and the anonymous
referee for an excellent and thorough report that improved the paper.
Support
from NASA grant NNX15AK14G, STScI grant GO-13681,
and NSF grants AST-1313280, AST-1515084, AST-1518294,
AST-1616598, and AST-1616710 is gratefully acknowledged.
C.~C.\ 
acknowledges support from the Packard Foundation.
A.~V.\ is supported by an NSF Graduate Research Fellowship.
Based on data obtained with the
W.~M.~Keck Observatory, on
Maunakea, Hawaii.
The authors wish to recognize and acknowledge the very significant
cultural role and reverence that the summit of Maunakea has always
had within the indigenous Hawaiian community.  We are most fortunate
to have the opportunity to conduct observations from this mountain.
\end{acknowledgements}

\begin{appendix}

\section{A.\ Fit Parameters}

The models we use have 36 free parameters. Several key ones are
described in \S\,3.2; here we provide a complete list of all model
parameters that are varied in the MCMC fit. The number between
square brackets is a running tally of the number of parameters.\vspace{0.2cm}\\
{\em Kinematics:} radial velocity $v$ [1] and velocity dispersion $\sigma$
[2].
\vspace{0.2cm}\\
{\em Star formation history:} two
stellar populations, with ages $\tau_1$ and $\tau_2$ [3,4] and the mass
ratio of the two populations [5].
\vspace{0.2cm}\\
{\em Metallicity:} total metallicity [Z/H] [6], and the individual
elements Fe, O, C, N, Na, Mg, Si, K, Ca, Ti, V, Cr, Mn, Co, Ni, Sr,
and Ba [7--23].
\vspace{0.2cm}\\
{\em Initial mass function:} logarithmic slopes in the mass
ranges
$0.08<M/{\rm M}_{\odot}<0.5$ ($x_1$)
and $0.5<M/{\rm M}_{\odot}<1.0$ ($x_2$) [24,25]. 
\vspace{0.2cm}\\
{\em Hot star component:} temperature ($T_{\rm eff,hot}$) and weight of
the component [26,27].
{\em Emission lines:} velocity dispersion of emission lines $\sigma_{\rm emi}$
[26], line fluxes of H, [O\,{\sc iii}], [S\,{\sc ii}], [N\,{\sc i}],
and [N\,{\sc ii}] [28--33].
\vspace{0.2cm}\\
{\em Atmospheric transmission:} normalization of atmospheric transmission
function (to allow for residual telluric absorption) [34].
{\em Error normalization:} correction applied to the observational
uncertainties, of the form $(\Delta f)_{\rm corr} = {\rm jitter}_{\rm all}
\times (\Delta f)_{\rm org} + {\rm jitter}_{\rm sky} \times f_{\rm sky}$,
with $\Delta f$ the uncertainties in the galaxy spectrum
and $f_{\rm sky}$ a template of the sky spectrum [35,36].

\section{B.\ Analysis of Systematic Residuals}

In \S\,3.2 we showed that the residuals from the model fits are strongly
correlated, in the sense that the residuals for
different galaxies (and
for different radial bins within galaxies) are very similar.
The green line in the top panels of Fig.\ 8 shows the median residual for
all radial bins in all six galaxies, and the bottom panels of that Figure
show the remaining systematic variation after subtracting this green line.
Here we quantify this variation, and compare the amplitude of the
residuals for subsets of the spectra to the strength of the IMF signal
that we are aiming to measure.

\begin{figure*}[htbp]
  \begin{center}
  \includegraphics[width=0.65\linewidth]{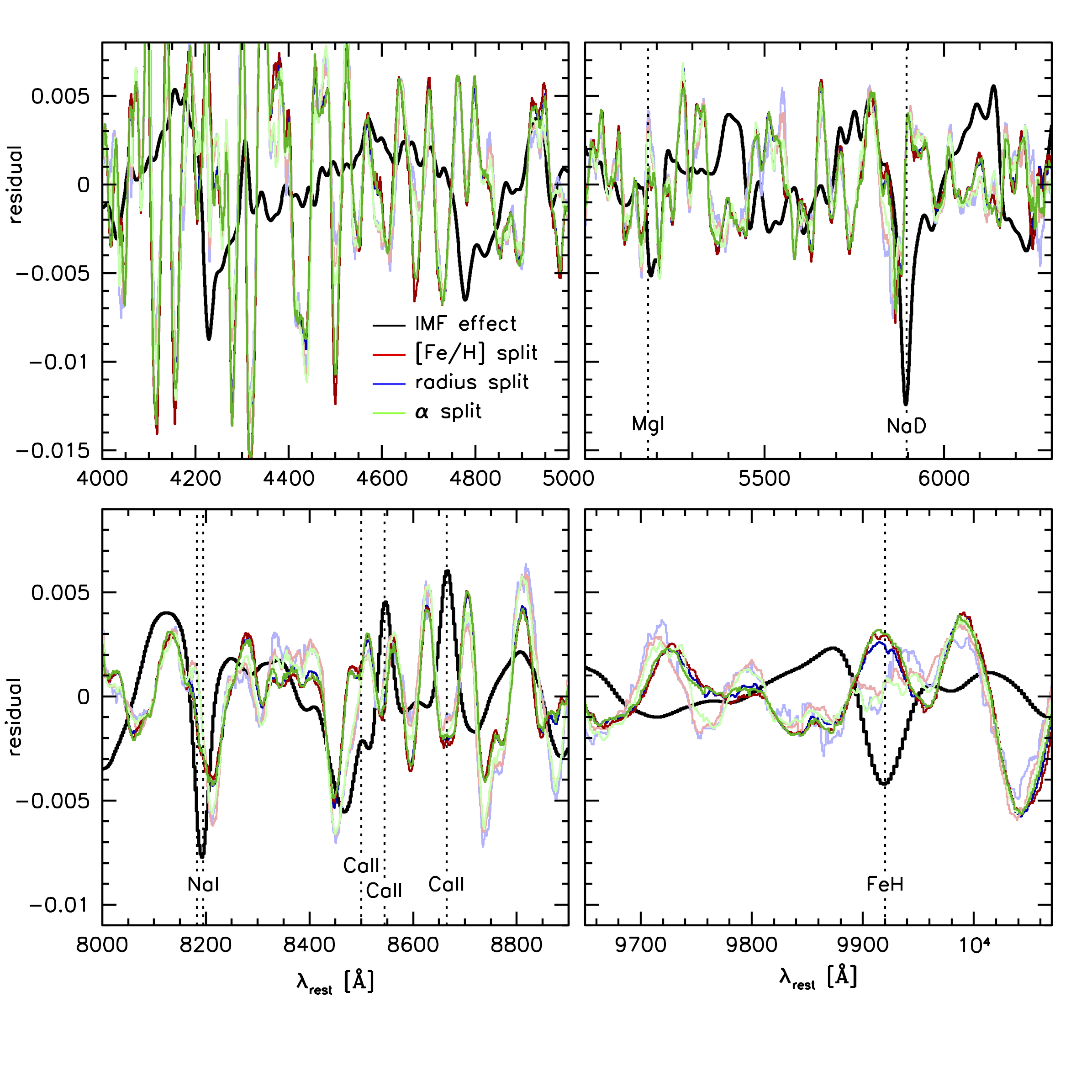}
  \end{center}
\vspace{-1.3cm}
    \caption{
Systematic median residuals from the fits, split by spectral properties.
Dark and light red curves show the residuals for [Fe/H]$\geq -0.1$
and [Fe/H]$<-0.1$ respectively. Dark and light blue curves are for
$R<0.2R_{\rm e}$ and $R\geq 0.2R_{\rm e}$. Dark and light green
curves are for IMF mismatch parameter $\alpha\geq 2$ and $\alpha<2$.
The residuals are very similar for all these subsamples.
The black curve shows the signal of variation in the IMF,
for $\Delta \alpha \sim 2.5$. 
}
\label{res_split.fig}
\end{figure*}

\begin{figure*}[htbp]
  \begin{center}
  \includegraphics[width=0.65\linewidth]{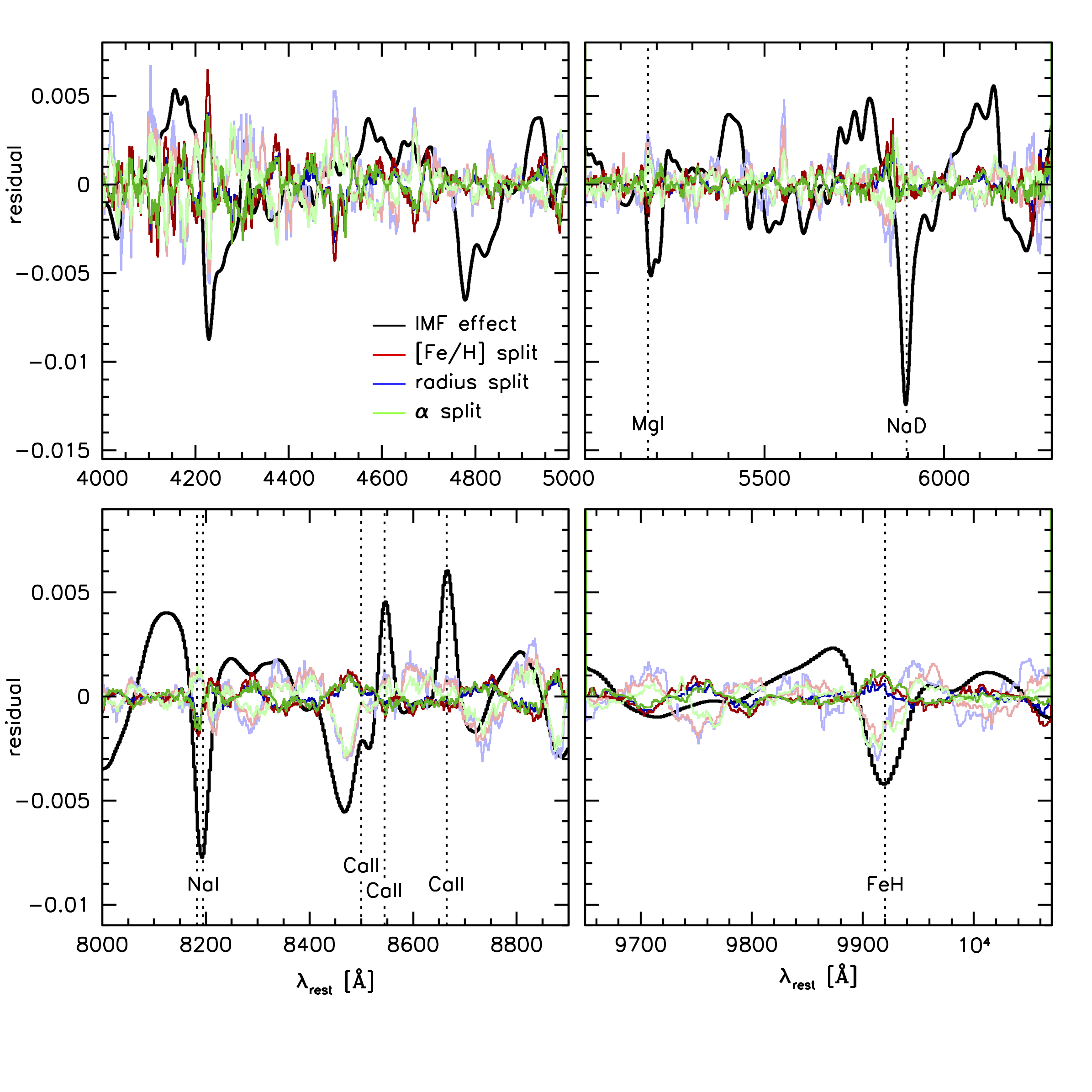}
  \end{center}
\vspace{-1.3cm}
    \caption{
Same as Fig.\ \ref{res_split.fig}, but now after subtracting the
median residual of all spectra (the green curve in Fig.\ 8 in the
main text). Even though the systematic residuals are of a similar
level as the signal of IMF variation,
the {\em variation} in the residuals as a function of [Fe/H],
$R$, and $\alpha$ is much smaller.
}
\label{res_split2.fig}
\end{figure*}

We split the 92 spectra (six galaxies, and an average of $\sim 15$
radial bins per galaxy) into two approximately equal-sized samples, based
on three criteria. First, we divide the spectra according to their
metallicity,  separately considering spectra with [Fe/H]$<-0.1$ and
those with [Fe/H]$\geq -0.1$. The median systematic residuals for these
two samples are shown with red lines in Fig.\ \ref{res_split.fig},
with dark red for the low metallicity sample and light red for the
high metallicity sample. In Fig.\ \ref{res_split2.fig} we show the
same curves after subtracting the median residual for the full sample
(the green line in Fig.\ 8 in the main text). We find
that the systematic residuals depend weakly on metallicity.
The rms of the systematic residual is $\approx 0.0023$ (0.23\,\%) in the
region $5000\,$\AA$<\lambda<6300$\,\AA\  for
both metallicity bins, and after subtracting the median residual of
the full sample the remaining rms is only $0.0007$ and $0.0006$ for
the two bins. The results are similar for the other spectral regions.
We also split the sample by radius (blue curves; dark blue is for
$R<0.2R_{\rm e}$ and light blue is for $R\geq 0.2R_{\rm e}$)
and by the IMF mismatch parameter (green; dark green is $\alpha\geq
2$ and light green is $\alpha<2$). These binnings produce very similar
curves as the metallicity bins, which is not surprising given the
strong correlations between radius, [Fe/H], and $\alpha$.

Importantly, these remaining residuals are weaker than the amplitude
of IMF variations. The black lines in Figs.\ \ref{res_split.fig} and
Fig.\ \ref{res_split2.fig} show the spectral response due to a change
in the IMF (see also, e.g., Conroy \& van Dokkum 2012a). The curves
were created by dividing a model spectrum for a stellar population with
an $x=3.0$ IMF by a model with an $x=2.3$ IMF, and subtracting the
continuum.\footnote{Both models have
an age of 13.5\,Gyr and Solar metallicity.} The difference in IMF
mismatch parameter between the two models is $\Delta \alpha \approx
2.5$. Even though the systematic residuals have a similar amplitude as
the IMF effect (see Fig.\ \ref{res_split.fig}), the {\em variation}
in the systematic residuals is a factor of $\sim 4$ smaller than the
IMF effect (Fig.\ \ref{res_split2.fig}). Moreover, with the exception of a
feature near the \feh\ band, there is no evidence for systematic variation
in the residuals at the locations of the ``classic'' IMF features
(\nad, \nai, and the Calcium triplet). The feature near FeH may
be caused by imperfect modeling of a strong TiO band, or simply
caused by noise: the remaining variations in the red part of the
spectrum are only a factor of $\sim 1.5$
higher than the expectation from photon noise.

We end by reiterating that the $\pm 0.2$\,\% systematic residuals
shown in Fig.\ \ref{res_split.fig} are likely due to deficiencies in the
stellar population synthesis models (see \S\,3.2). Although it is desirable to
improve the models and reduce these residuals, we note that they
do not adversely influence the formal errors on the derived parameters:
one of the parameters in the fit (see Appendix A) multiplies the formal errors
to ensure that the $\chi^2$ is acceptable.

\section{C.\ The Wing-Ford Band}

In the main text we do not measure or analyze the strength of individual
absorption features. Instead, the spectra are fit in their entirety
(``full spectrum fitting''; see, e.g., Conroy \& van Dokkum 2012b).
For completeness, and for comparison to other studies,
we here discuss the strength of the \feh\ Wing-Ford
band for the extracted spectra.
The FeH molecular band is very strong in low mass stars and
absent in giants (Wing \& Ford 1969; Schiavon 1997), and
of all
individual absorption features in the optical it
may be expected to show the strongest correlation with the IMF.
However, as with all individual features, the interpretation is
not straightforward.
As discussed in
Conroy \& van Dokkum (2012a) and {La Barbera} {et~al.} (2016)
FeH may not show an IMF-dependence if there are
counter-acting variations in metallicity or other parameters. Furthermore,
there is a contaminating  TiO band head at 9900\,\AA\
($\delta\,R_1(23)\,2-3$; Valenti et al.\ 1998).

\begin{figure*}[htbp]
  \begin{center}
  \includegraphics[width=0.7\linewidth]{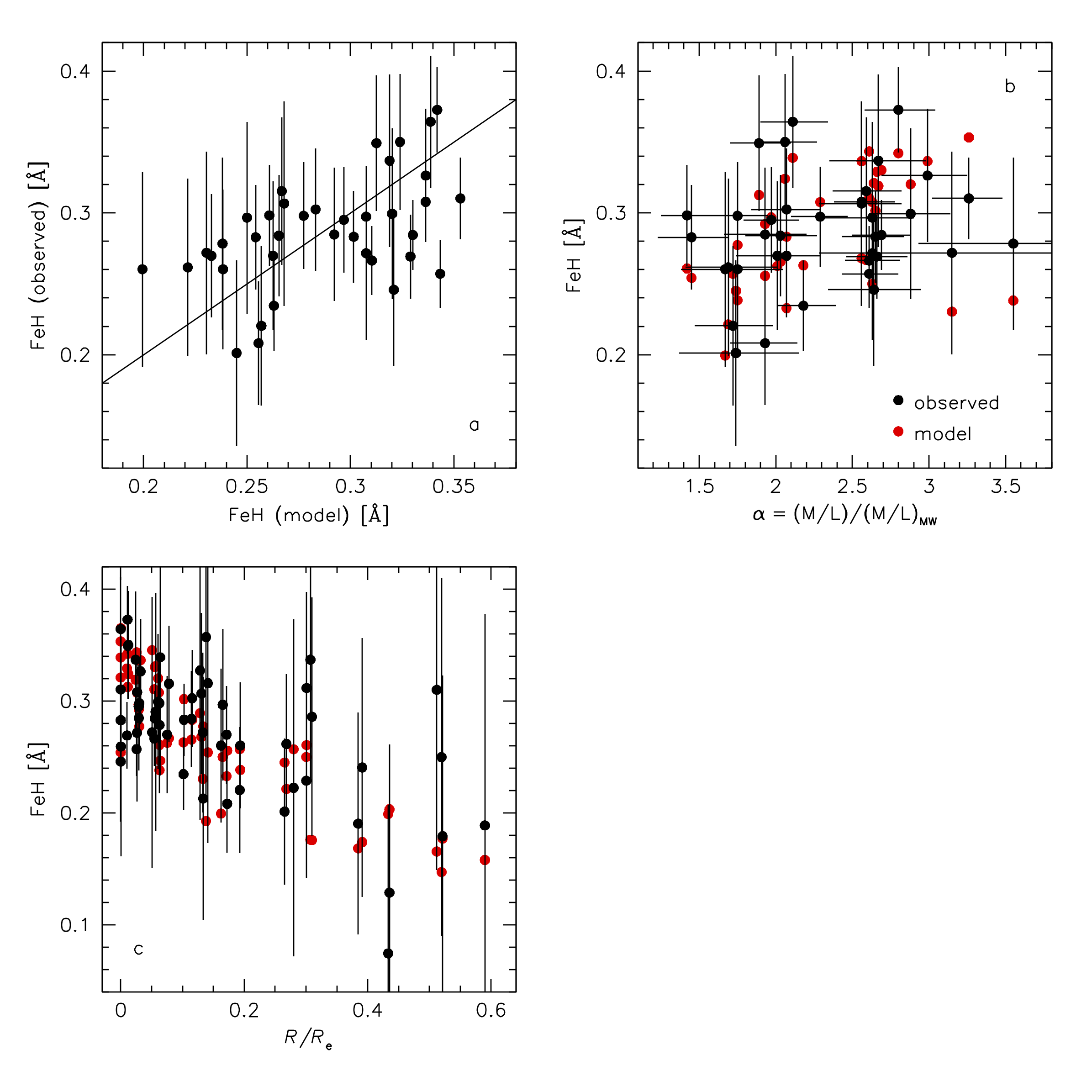}
  \end{center}
\vspace{-0.8cm}
    \caption{
{\it (a)}
Strength of the observed \feh\ absorption versus that in the best-fitting
model, only showing points with errors $<0.08$\,\AA.
The solid line is the line of equality.
The models reproduce the observed line strengths within the errors.
{\it (b)} Relation between the FeH index and the IMF. There is no strong
correlation in this sample of high quality spectra. This is not just
true for the data but also for the model spectra, reflecting the fact
that there are other parameters that influence the FeH absorption.
{\it (c)} Relation between FeH index and radius, now showing points with
errors $<0.02$\,\AA. There is a strong relation, likely driven by a
combination of the IMF and metallicity.
}
\label{wf.fig}
\end{figure*}

With these caveats in mind we measure the strength of FeH
at all radii for all six galaxies, using
the definition of van Dokkum \& Conroy (2010).\footnote{Note that the
van Dokkum \& Conroy (2010) index definition is different from that
of Conroy \& van Dokkum (2012a): the latter is appropriate for 
spectra at high resolution, whereas the former was chosen for
spectra at low resolution.}
The measurements
are performed
on the de-redshifted spectra, smoothed to a common
resolution of $\sigma=450$\,\kms.
We also measure FeH in the best-fitting models, at the
same resolution.
In Fig.\ \ref{wf.fig}a the observed FeH index is compared to that
in the best-fitting model. The units are \AA; they
can be converted to the average absorbed
continuum fraction within the band (as
shown in  van Dokkum \& Conroy 2012b) by dividing them by the
width of the band (20\,\AA).
Only points with errors $<0.08$\,\AA\ are shown.
The rms scatter around the line of equality
is 0.044\,\AA, and consistent with the measurement errors. 

The FeH index is compared to the IMF ``mismatch'' parameter $\alpha$
in Fig.\ \ref{wf.fig}b. There is no correlation, even though $\alpha$
varies from 1.5 to 3.5 in this sample.\footnote{We do find
$\alpha<1.5$ for some galaxies and radial ranges,
as shown in the main text,
but for those spectra the errors in FeH exceed the
0.08\,\AA\ threshold (as these tend to be at large
radii where the S/N ratio is lower).}
Crucially, there is no strong correlation between the {\em model} FeH index
and the IMF either in this subsample of high quality spectra. The
rms of the model points (red) is 0.040\,\AA, and it is barely reduced
(to 0.036\,\AA) after subtracting the best-fitting linear relation.
Some of the weakest FeH bands are found among models that have
$\alpha>3$, i.e., very bottom-heavy IMFs.
This result reflects the fact that the Wing-Ford band alone is
not sufficient for measuring the IMF, as discussed above.

Finally, we show the relation between FeH and radius in Fig.\
\ref{wf.fig}c. Here we also show points with larger errorbars (up to
0.2\,\AA), to extend the radial range. There is a strong anti-correlation,
with the FeH band decreasing with radius in both the models and the data.
This trend is likely due to a combination of the IMF and metallicity,
as [Fe/H] decreases with radius.
These panels highlight
the importance of techniques that take all available information in
the spectrum into account.

\end{appendix}


\begin{references}

\reference{} {Barro}, G., {Faber}, S.~M., {P{\'e}rez-Gonz{\'a}lez}, P.~G., {Koo}, D.~C.,  {Williams}, C.~C., {Kocevski}, D.~D., {Trump}, J.~R., {Mozena}, M., {et al.} 2013, \apj, 765, 104

\reference{} {Barro}, G., {Kriek}, M., {P{\'e}rez-Gonz{\'a}lez}, P.~G., {Trump}, J.~R.,  {Koo}, D.~C., {Faber}, S.~M., {Dekel}, A., {Primack}, J.~R., {et al.} 2016, \apjl, 827, L32

\reference{} {Belli}, S., {Newman}, A.~B., \& {Ellis}, R.~S. 2014, \apj,
783, 117

\reference{} {Belli}, S., {Newman}, A.~B., \& {Ellis}, R.~S. 2016, \apj,
submitted (arXiv:1608.00608)

\reference{} {Bensby}, T., {Feltzing}, S., \& {Oey}, M.~S. 2014, \aa, 562, 71

\reference{} {Bernardi}, M., {Meert}, A., {Vikram}, V.,
{Huertas-Company}, M., {Mei}, S., {Shankar}, F., \& {Sheth}, R.~K. 2014,
\mnras, 443, 874

\reference{} {Bezanson}, R., {van Dokkum}, P.~G., {Tal}, T., {Marchesini}, D., {Kriek}, M.,  {Franx}, M., \& {Coppi}, P. 2009, \apj, 697, 1290

\reference{} {Boroson}, T.~A. \& {Thompson}, I.~B. 1991, \aj, 101, 111

\reference{} {Brough}, S., {Forbes}, D.~A., {Kilborn}, V.~A., {Couch}, W., \& {Colless}, M.  2006, \mnras, 369, 1351

\reference{} {Cappellari}, M., {Emsellem}, E., {Bacon}, R., {Bureau}, M., {Davies}, R.~L.,  {de Zeeuw}, P.~T., {Falc{\'o}n-Barroso}, J., {Krajnovi{\'c}}, D., {et al.} 2007, \mnras, 379, 418

\reference{} {Cappellari}, M., {McDermid}, R.~M., {Alatalo}, K., {Blitz}, L., {Bois}, M.,  {Bournaud}, F., {Bureau}, M., {Crocker}, A.~F., {et al.} 2012,
Nature, 484, 485

\reference{} {Cappellari}, M., {McDermid}, R.~M., {Alatalo}, K., {Blitz}, L., {Bois}, M.,  {Bournaud}, F., {Bureau}, M., {Crocker}, A.~F., {et al.} 2013, \mnras, 432, 1862

\reference{} {Cenarro}, A.~J., {Gorgas}, J., {Vazdekis}, A., {Cardiel}, N., \& {Peletier},  R.~F. 2003, \mnras, 339, L12

\reference{} {Chabrier}, G. 2003, \pasp, 115, 763

\reference{} {Chabrier}, G., {Hennebelle}, P., \& {Charlot}, S. 2014, \apj, 796, 75

\reference{} {Choi}, J., {Dotter}, A., {Conroy}, C., {Cantiello}, M., {Paxton}, B., \&  {Johnson}, B.~D. 2016, \apj, 823, 102

\reference{} {Conroy}, C., {Dutton}, A.~A., {Graves}, G.~J., {Mendel}, J.~T., \& {van  Dokkum}, P.~G. 2013, \apjl, 776, L26

\reference{} {Conroy}, C., {Graves}, G.~J., \& {van Dokkum}, P.~G. 2014, \apj, 780, 33

\reference{} {Conroy}, C. \& {van Dokkum}, P. 2012a, \apj, 747, 69

\reference{} {Conroy}, C. \& {van Dokkum}, P.~G. 2012b, \apj, 760, 71

\reference{} {Conroy}, C., {van Dokkum}, P., \& {Villaume}, A. 2017,
\apj, submitted (arXiv:1612.00013)

\reference{} {Davis}, T.~A. \& {McDermid}, R.~M. 2017, \mnras, 464, 453

\reference{} {Dutton}, A.~A., {Mendel}, J.~T., \& {Simard}, L. 2012, \mnras, 422, L33

\reference{} {Fang}, J.~J., {Faber}, S.~M., {Koo}, D.~C., \& {Dekel}, A. 2013, \apj, 776, 63

\reference{} {Foreman-Mackey}, D., {Hogg}, D.~W., {Lang}, D., \& {Goodman}, J. 2013, \pasp,  125, 306

\reference{} {Geha}, M., {Brown}, T.~M., {Tumlinson}, J., {Kalirai}, J.~S., {Simon}, J.~D.,  {Kirby}, E.~N., {VandenBerg}, D.~A., {Mu{\~n}oz}, R.~R., {et al.} 2013, \apj, 771, 29

\reference{} {Graves}, G.~J., {Faber}, S.~M., {Schiavon}, R.~P., \& {Yan}, R. 2007, \apj,  671, 243

\reference{} {Greene}, J.~E., {Janish}, R., {Ma}, C.-P., {McConnell}, N.~J., {Blakeslee},  J.~P., {Thomas}, J., \& {Murphy}, J.~D. 2015, \apj, 807, 11

\reference{} {Hopkins}, P.~F. 2013, \mnras, 433, 170

\reference{} {Kobayashi}, C., {Umeda}, H., {Nomoto}, K., {Tominaga}, N., \&
{Ohkubo}, T. 2006, \apj, 653, 1145

\reference{} {Kelson}, D.~D., {Zabludoff}, A.~I., {Williams}, K.~A., {Trager}, S.~C.,  {Mulchaey}, J.~S., \& {Bolte}, M. 2002, \apj, 576, 720

\reference{} {Krajnovi{\'c}}, D., {Alatalo}, K., {Blitz}, L., {Bois}, M., {Bournaud}, F.,  {Bureau}, M., {Cappellari}, M., {Davies}, R.~L., {et al.} 2013, \mnras,  432, 1768

\reference{} {Kroupa}, P. 2001, \mnras, 322, 231

\reference{} {Kuntschner}, H., {Emsellem}, E., {Bacon}, R., {Cappellari}, M., {Davies},  R.~L., {de Zeeuw}, P.~T., {Falc{\'o}n-Barroso}, J., {Krajnovi{\'c}}, D., {et al.} 2010, \mnras, 408, 97

\reference{} {La Barbera}, F., {Ferreras}, I., \& {Vazdekis}, A. 2015, \mnras, 449, L137

\reference{} {La Barbera}, F., {Ferreras}, I., {Vazdekis}, A., {de la Rosa}, I.~G., {de  Carvalho}, R.~R., {Trevisan}, M., {Falc{\'o}n-Barroso}, J., \&  {Ricciardelli}, E. 2013, \mnras, 433, 3017

\reference{} {La Barbera}, F., {Vazdekis}, A., {Ferreras}, I., {Pasquali}, A., {Cappellari},  M., {Mart{\'{\i}}n-Navarro}, I., {Sch{\"o}nebeck}, F., \&  {Falc{\'o}n-Barroso}, J. 2016, \mnras, 457, 1468

\reference{} {L{\"a}sker}, R., {van den Bosch}, R.~C.~E., {van de Ven}, G.,
{Ferreras}, I., {La Barbera}, F., {Vazdekis}, A., \& {Falc{\'o}n-Barroso}, J.
2013, \mnras, 434, L31

\reference{} {Leier}, D., {Ferreras}, I., {Saha}, P., {Charlot}, S., {Bruzual}, G., \& {La  Barbera}, F. 2016, \mnras, 459, 3677

\reference{} {Li}, Z.-Y., {Ho}, L.~C., {Barth}, A.~J., \& {Peng}, C.~Y. 2011, \apjs, 197, 22

\reference{} {Lyubenova}, M., {Mart{\'{\i}}n-Navarro}, I., {van de Ven}, G.,  {Falc{\'o}n-Barroso}, J., {Galbany}, L., {Gallazzi}, A.,  {Garc{\'{\i}}a-Benito}, R., {Gonz{\'a}lez Delgado}, R., {et al.} 2016, \mnras, 463, 3220

\reference{} {Mann}, A.~W., {Feiden}, G.~A., {Gaidos}, E., {Boyajian}, T., \& {von Braun},  K. 2015, \apj, 804, 64

\reference{} {Mart{\'{\i}}n-Navarro}, I., {Barbera}, F.~L., {Vazdekis}, A.,  {Falc{\'o}n-Barroso}, J., \& {Ferreras}, I. 2015a, \mnras, 447,  1033

\reference{} {Mart{\'{\i}}n-Navarro}, I., {La Barbera}, F., {Vazdekis}, A.,  {Ferr{\'e}-Mateu}, A., {Trujillo}, I., \& {Beasley}, M.~A.  2015b, \mnras, 451, 1081

\reference{} {Mart{\'{\i}}n-Navarro}, I., {Vazdekis}, A., {La Barbera}, F.,
{Falc{\'o}n-Barroso}, J., {Lyubenova}, M., {van de Ven}, G.,
{Ferreras}, I., {S{\'a}nchez}, {et al.} 2015c, \apj, 806, L31

\reference{} {McConnell}, N.~J., {Lu}, J.~R., \& {Mann}, A.~W. 2016, \apj, 821, 39

\reference{} {Mehlert}, D., {Thomas}, D., {Saglia}, R.~P., {Bender}, R., \& {Wegner}, G.  2003, \aap, 407, 423

\reference{} {Milone}, A.~D.~C. and {Sansom}, A.~E., \&
{S{\'a}nchez-Bl{\'a}zquez}, P. 2011, \mnras, 414, 1227

\reference{} {Nelson}, E., {van Dokkum}, P., {Franx}, M., {Brammer}, G., {Momcheva}, I.,  {Schreiber}, N.~F., {da Cunha}, E., {Tacconi}, L., {et al.} 2014, \nat, 513, 394

\reference{} {Newman}, A.~B., {Belli}, S., {Ellis}, R.~S.\ 2015, \apj,
813, L7

\reference{} {Newman}, A.~B., {Smith}, R.~J., {Conroy}, C., {Villaume}, A.,
\& {van Dokkum}, P. 2017, ApJ, submitted (arXiv:1612.00065)

\reference{} {Oke}, J.~B., {Cohen}, J.~G., {Carr}, M., {Cromer}, J., {Dingizian}, A.,  {Harris}, F.~H., {Labrecque}, S., {Lucinio}, R., {et al.} 1995, \pasp, 107, 375

\reference{} {Oser}, L., {Ostriker}, J.~P., {Naab}, T., {Johansson}, P.~H., \& {Burkert}, A.  2010, \apj, 725, 2312

\reference{} {Peacock}, M.~B., {Zepf}, S.~E., {Maccarone}, T.~J., {Kundu}, A., {Gonzalez},  A.~H., {Lehmer}, B.~D., \& {Maraston}, C. 2014, \apj, 784, 162

\reference{} {Posacki}, S., {Cappellari}, M., {Treu}, T., {Pellegrini}, S., \& {Ciotti}, L.  2015, \mnras, 446, 493

\reference{} {Rockosi}, C., {Stover}, R., {Kibrick}, R., {Lockwood}, C., {Peck}, M.,  {Cowley}, D., {Bolte}, M., {Adkins}, S., {et al.} 2010, in  Society of Photo-Optical Instrumentation Engineers (SPIE) Conference Series,  Vol. 7735, Society of Photo-Optical Instrumentation Engineers (SPIE)  Conference Series

\reference{} {Romanowsky}, A.~J., {Strader}, J., {Spitler}, L.~R., {Johnson}, R., {Brodie},  J.~P., {Forbes}, D.~A., \& {Ponman}, T. 2009, \aj, 137, 4956

\reference{} {Saglia}, R.~P., {Maraston}, C., {Thomas}, D., {Bender}, R., \& {Colless}, M.  2002, \apjl, 579, L13

\reference{} {Salpeter}, E.~E. 1955, \apj, 121, 161

\reference{} {S{\'a}nchez-Bl{\'a}zquez}, P., {Gorgas}, J., {Cardiel}, N., \& {Gonz{\'a}lez},  J.~J. 2006, \aap, 457, 787

\reference{} {Schiavon}, R.~P. 2007, \apjs, 171, 146

\reference{} {Schiavon}, R.~P., {Barbuy}, B., \& {Singh}, P.~D. 1997,
\apj, 484, 499

\reference{} {Schwartz}, C.~M. \& {Martin}, C.~L. 2004, \apj, 610, 201

\reference{} {Sersic}, J.~L. 1968, {Atlas de galaxias australes} (Cordoba, Argentina:  Observatorio Astronomico, 1968)

\reference{} {Smith}, R.~J. 2014, \mnras, 443, L69

\reference{} {Smith}, R.~J., {Lucey}, J.~R., \& {Carter}, D. 2012, \mnras, 426, 2994

\reference{} {Smith}, R.~J., {Lucey}, J.~R., \& {Conroy}, C. 2015, \mnras, 449, 3441

\reference{} {Smith}, R.~M., {Mart{\'{\i}}nez}, V.~J., {Fern{\'a}ndez-Soto}, A.,  {Ballesteros}, F.~J., \& {Ortiz-Gil}, A. 2008, \apj, 679, 420

\reference{} {Sonnenfeld}, A., {Nipoti}, C., \& {Treu}, T. 2016, \mnras,
submitted (arXiv:1607.01394)

\reference{} {Sonnenfeld}, A., {Treu}, T., {Marshall}, P.~J., {Suyu}, S.~H., {Gavazzi}, R.,  {Auger}, M.~W., \& {Nipoti}, C. 2015, \apj, 800, 94

\reference{} {Spiniello}, C., {Barnab{\`e}}, M., {Koopmans}, L.~V.~E., \& {Trager}, S.~C.  2015a, \mnras, 452, L21


\reference{} {Spiniello}, C., {Napolitano}, N.~R., {Pota}, V.,
{Romanowsky}, A.~J., {Tortora}, C., {Covone}, G., \& {Capaccioli},
M. 2015b, \mnras, 452, 99

\reference{} {Spiniello}, C., {Koopmans}, L.~V.~E., {Trager}, S.~C., {Czoske}, O., \&  {Treu}, T. 2011, \mnras, 417, 3000

\reference{} {Spiniello}, C., {Trager}, S.~C., {Koopmans}, L.~V.~E., \& {Chen}, Y. 2012, \apj, 753, L32

\reference{} {Spolaor}, M., {Forbes}, D.~A., {Hau}, G.~K.~T., {Proctor}, R.~N., \& {Brough},  S. 2008a, \mnras, 385, 667

\reference{} {Spolaor}, M., {Forbes}, D.~A., {Proctor}, R.~N., {Hau}, G.~K.~T., \& {Brough},  S. 2008b, \mnras, 385, 675

\reference{} {Tang}, B. \& {Worthey}, G. 2015, \mnras, 453, 4431

\reference{} {Thomas}, J., {Ma}, C.-P., {McConnell}, N.~J., {Greene}, J.~E., {Blakeslee},  J.~P., \& {Janish}, R. 2016, \nat, 532, 340

\reference{} {Thomas}, J., {Saglia}, R.~P., {Bender}, R., {Thomas}, D., {Gebhardt}, K.,  {Magorrian}, J., {Corsini}, E.~M., {Wegner}, G., {et al.} 2011, \mnras,  415, 545

\reference{} {Tortora}, C., {Napolitano}, N.~R., {Romanowsky}, A.~J., {Jetzer}, P.,  {Cardone}, V.~F., \& {Capaccioli}, M. 2011, \mnras, 418, 1557

\reference{} {Trager}, S.~C., {Faber}, S.~M., {Worthey}, G., \& {Gonz{\'a}lez}, J.~J. 2000,  \aj, 119, 1645

\reference{} {Treu}, T., {Auger}, M.~W., {Koopmans}, L.~V.~E., {Gavazzi}, R., {Marshall},  P.~J., \& {Bolton}, A.~S. 2010, \apj, 709, 1195

\reference{} {Valenti}, J.~A., {Piskunov}, N., \& {Johns-Krull}, C.~M. 1998,
\apj, 498, 851

\reference{} {van de Sande}, J., {Kriek}, M., {Franx}, M., {van Dokkum}, P.~G., {Bezanson},  R., {Bouwens}, R.~J., {Quadri}, R.~F., {Rix}, H.-W., {et al.} 2013, \apj, 771, 85

\reference{} {van der Marel}, R.~P. \& {Franx}, M. 1993, \apj, 407, 525

\reference{} {van Dokkum}, P.~G., {Bezanson}, R., {van der Wel}, A., {Nelson}, E.~J.,  {Momcheva}, I., {Skelton}, R.~E., {Whitaker}, K.~E., {Brammer}, G., {et al.} 2014, \apj, 791, 45

\reference{} {van Dokkum}, P.~G. \& {Conroy}, C. 2010, \nat, 468, 940

\reference{} ---. 2011, \apjl, 735, L13

\reference{} ---. 2012, \apj, 760, 70

\reference{} {van Dokkum}, P.~G. \& {Franx}, M. 1995, \aj, 110, 2027

\reference{} {van Dokkum}, P.~G., {Nelson}, E.~J., {Franx}, M., {Oesch}, P., {Momcheva}, I.,  {Brammer}, G., {F{\"o}rster Schreiber}, N.~M., {Skelton}, R.~E., {et al.} 2015, \apj, 813, 23

\reference{} {Villaume}, A., et al.\ 2016, \apj, submitted

\reference{} {Wing}, R.~F. \& {Ford}, Jr., W.~K. 1969, \pasp, 81, 527

\reference{} {Worthey}, G., {Ingermann}, B.~A., \& {Serven}, J. 2011, \apj, 729, 148

\reference{} {Zieleniewski}, S., {Houghton}, R.~C.~W., {Thatte}, N., {Davies}, R.~L., \&  {Vaughan}, S.~P. 2016, \mnras, in press (arXiv:1611.01095)

\reference{} {Zolotov}, A., {Dekel}, A., {Mandelker}, N., {Tweed}, D., {Inoue}, S.,  {DeGraf}, C., {Ceverino}, D., {Primack}, J.~R., {et al.} 2015, \mnras, 450, 2327

\end{references}
\end{document}